\documentclass{usm-thesis}

\usepackage{lipsum}
\usepackage[english]{babel}
\usepackage{slashed}
\usepackage{placeins}
\usepackage{listings}
\usepackage{graphicx}
\DeclareUnicodeCharacter{2212}{-}

\definethesis{Probing Left-handed Heavy Neutral Leptons on the Massive Vector Doublet Model 
}{Paulo Andr\'es Areyuna Calabrese   
}

\begin{document}

\thesistitlepage{Advisors: Alfonso R. Zerwekh\\
\quad\quad \quad \quad \quad Jilberto Zamora Sa\'a 
}{Thesis submitted in partial fulfillment for the degree of Magister en Ciencias, menci\'on F\'isica, from Universidad T\'ecnica Federico Santa Mar\'ia. 
}{December, 2022 
}

\frontmatter

\committee{
  {Alfonso Zerwekh}{USM},
  {Jilberto Zamora}{UNAB},
  {Antonio Cárcamo}{USM},
  {Giovanna Cottin}{UAI}
}

\Dedication{to my family}

\CC[by-sa]
\newpage
\textbf{Acknowledgements}
\newline

First of all, I would like to thank my advisors, Alfonso Zerwekh and Jilberto Zamora, for their guidance and patience during these two years. Many times, they were advisors in a more general sense than the academical.

On the other side, I am very grateful for the unconditional love and support from my family: my parents,  Ariel and Rossana, and my siblings, Johara and Ernesto. I owe  every achievement I have gotten to my family and this is not the exception.

Of course, there is place in these words to my friends and colleagues from the university, special mentions to the \emph{Dark Matter Hunters}: Gonzalo Benítez, Patricio Escalona and Sebastián Acevedo.

Finally, I thank SAPHIR Millenium Institute. This work was funded by ANID - Millennium Program - ICN2019\_044. Also, I would like to thank to the DGIIP-UTFSM for founding during the development of this work.

\tableofcontents

\mainmatter

\chapter{The Standard Model of Particle Physics}
At the fundamental level, the universe has four interactions, which are the strong force, the weak force, the electromagnetic force and gravity. The standard model of particle physics (SM) is a theory that describes all the fundamental forces except gravity. In this chapter we review the fundamental concepts behind the standard model, giving special attention to the electroweak sector. To do so, we follow Ref. \cite{langacker}.

\section{The key ingredients}
The standard model, at the most basic level, is a theory that describes how particles interact. 
These interactions are described by means of quantum field theory.
This theory is built over the idea of the symmetry. Under this framework, the interactions are a natural consequence of symmetry. Since the SM symmetry is expected to be local (i.e. respected in every point of space and time), the fundamental forces manifest themselves via the exchange of spin 1 bosonic particles. In this section, we show how to describe the concepts presented here in a mathematically consistent way.
\subsection{The gauge group}
The gauge symmetry group of the SM is the following:
\begin{equation}\label{smgroup}
    G_{SM}=SU(3)_c\times SU(2)_L\times U(1)_Y.
\end{equation}
The subscripts in the right side of eq. \eqref{smgroup} have no mathematical meaning, but they are included in the literature because they are related with the physical meaning of the corresponding groups. The SM requires local gauge invariance under this group.  This condition forces the introduction of a vector field for every generator of the group. As we will see, these vector fields are the corresponding force carriers of the fundamental interactions.
A detailed description of the gauge group and the corresponding parameters can be seen in Table \ref{gaugegroup}.
\begin{table}[!h]
    \centering
    \begin{tabular}{|c|c|c|c|}
    \hline
        Group & $SU(3)_c$ & $SU(2)_L$ &  $U(1)_Y$  \\
        \hline
        Number of Generators &  8    &  3  & 1     \\
        Gauge bosons &$G_\mu^i$ &  $W_\mu^i$    &  $B_{\mu}$\\
        Coupling constant  &   $g_s$     &  $g$  &  $g'$\\
        Group Charge &  color ($c$) & Weak isospin ($\overrightarrow{
        T}$)    &   Hypercharge ($Y=Q-T_3$)    \\
        Field Strength & $G_{\mu\nu}^i$& $W_{\mu\nu}^i$ & $B_{\mu\nu}$\\
        \hline
    \end{tabular}
    \caption{Group description for the Standard Model}
    \label{gaugegroup}
\end{table}
We recall the form of the field strength tensor $X_{\mu\nu}$ for a gauge field $A_\mu$:
\begin{equation}
        X_{\mu\nu}^i =\partial_\mu A_\nu^i-\partial_\nu A_\mu^i-g_A c_{jk}^i A_\mu^j A_\nu^k,
        \end{equation}
        where $g_a$ is the coupling constant and $c_{jk}^i$ are the structure constants of the gauge group. This tensor is used to define the kinetic term of the gauge bosons, which has the following form:
        \begin{equation} \mathcal{L}_{gauge}=-\frac{1}{4}X_{\mu\nu}^iX^{i\mu\nu }.
        \end{equation}

It's worth to mention the relationship between groups and forces:
The $SU(3)$ group represents the strong interactions, while the $SU(2)\times U(1)$ group stands for the electroweak interaction.

\subsection{The matter content}

The matter content of the SM can be classified in two groups: quarks and leptons. These particles are described by fermionic fields, which transform under the SM gauge group as shown in Table \ref{matter}. For simplicity, we considered the first generation of quarks and leptons.
We can see that the $SU(2)_L\times U(1)_Y$ group only transforms left-handed fermions in a non trivial way. We also note the presence of right-handed neutrinos, which are SM singlets and have no gauge interactions.
\begin{table}[!h]
    \centering
\begin{tabular}{|c|c|c|c|}
\hline
Field &  $SU(3)_c$&$ SU(2)_L$ & $U(1)_Y$ \\
    \hline
    $\begin{pmatrix}\nu_{eL}\\e_L^-\end{pmatrix}\equiv L$ & $1$&$2$&$-1/2$\\
    \hline
    $\nu_R$  & $1$ & $1$ &$0$ \\
    \hline
    $e_R$ & $1$  &$1$&$-1$\\
    \hline
        $\begin{pmatrix}u_L\\d_L\end{pmatrix}\equiv q$ & $3$&$2$&$1/6$\\
    \hline
    $u_R$  & $3$ & $1$ &$2/3$ \\
    \hline
    $d_R$ & $3$  &$1$&$-1/3$\\
    \hline

\end{tabular}
    \caption{Group representation of the matter content}
    \label{matter}
\end{table}

The interaction of gauge bosons and fermions is built in the minimal setting, which consists in the replacement of the derivative by the covariant derivative, defined as:
\begin{equation}
    D_{\mu}=\partial_\mu +i\sum_{k}g_kQ_kT_kV_\mu^k,
\end{equation}
where $g,Q,T,V$ stand for the coupling constant, the group charge, the generators and the associated boson, respectively. The minimal setting is necessary to fulfill the local gauge invariance. With these considerations, the kinetic terms for the fermion sector can be written as:
\begin{equation}
    \mathcal{L}_{fermion}=i\bar{L}\slashed{D}L+i\bar{\nu}_R\slashed{D}\nu_R+i\bar{e}_R\slashed{D}e_R+
    i\bar{q}\slashed{D}q
    +i\bar{u}_R\slashed{D}u_R+i\bar{d}_R\slashed{D}d_R.
\end{equation}
At this level, all particles are massless. The inclusion of mass terms would break gauge invariance. In order to include mass terms for fermions we need  to include the Higgs mechanism for Spontaneous Symmetry Breaking (SBB for short).
\subsection{The Higgs Sector}
\begin{figure}[!h]
    \centering
    \includegraphics[width=0.6\linewidth]{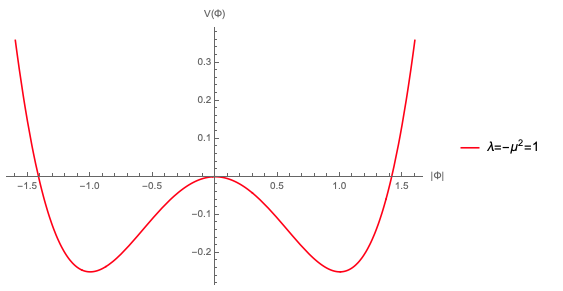}
    \caption{The Higgs potential as a function of $|\Phi|$}
    \label{higgspot}
\end{figure}

The introduction of masses in the theory can be cleverly done by adding the
Higgs doublet, which is defined as follows:
\begin{equation}
    \Phi=\begin{pmatrix}\phi^+\\\phi^0\end{pmatrix}=\begin{pmatrix}\frac{\phi_1+i\phi_2}{\sqrt{2}}\\\frac{\phi_3+i\phi_4}{\sqrt{2}}\end{pmatrix}.
\end{equation}
This doublet transforms as $(1,2,1/2)$ under the SM group. The key ingredient is the definition of the Higgs potential:
\begin{equation}
    V(\Phi)=\frac{\mu^2}{2}|\Phi|^2+\frac{\lambda}{4}|\Phi|^4,
\end{equation}
with $|\Phi|^2=\Phi^\dagger\Phi$. This potential is sketched in Figure \ref{higgspot}. The potential has two minimum points:
\begin{equation}
    |\Phi|_{min}=\pm\sqrt{\frac{-\mu^2}{\lambda}}.
\end{equation}

As we stated before, the SM is a quantum field theory. In order to be able to apply all the tools of quantum field theory, it's necessary to define the vacuum of the theory, which corresponds to the field configuration where the potential is minimum.
Unlike most of the fields, the Higgs doublet acquires a non vanishing vacuum expectation value (vev for short), which has the following form:
\begin{equation}
   \langle \Phi\rangle= \langle 0 |\Phi|0\rangle=\frac{1}{\sqrt{2}}\begin{pmatrix}0\\v\end{pmatrix},
\end{equation}
With $v=\sqrt{\frac{-\mu^2}{\lambda}}$. It is worth mentioning that there are infinite minimum points, constrained by the general conditions:
\begin{equation}
    \phi_1^2+\phi_2^2+\phi_3^2+\phi_4^2=2v^2,
\end{equation}
however, all the fields configurations are equivalent because it's possible to relate them by applying an appropiate $SU(2)_L$ transformation. The non zero vev induces the so called spontaneous symmetry breaking. To identify the broken symmetries, we need to study the effect of the group generators over the vev:
\begin{equation}
    \begin{split}
        &T_1\langle \Phi\rangle=\frac{\sigma_1}{2}\langle \Phi\rangle=\frac{1}{2}\begin{pmatrix}0&1\\1&0\end{pmatrix}\begin{pmatrix}0\\\frac{v}{\sqrt{2}}\end{pmatrix}=\frac{1}{2}\begin{pmatrix}\frac{v}{\sqrt{2}}\\0\end{pmatrix}\neq 0\\
         &T_2\langle \Phi\rangle=\frac{\sigma_2}{2}\langle \Phi\rangle=\frac{1}{2}\begin{pmatrix}0&-i\\i&0\end{pmatrix}\begin{pmatrix}0\\\frac{v}{\sqrt{2}}\end{pmatrix}=-\frac{i}{2}\begin{pmatrix}\frac{v}{\sqrt{2}}\\0\end{pmatrix}\neq 0\\
          &T_3\langle \Phi\rangle=\frac{\sigma_3}{2}\langle \Phi\rangle=\frac{1}{2}\begin{pmatrix}1&0\\0&-1\end{pmatrix}\begin{pmatrix}0\\\frac{v}{\sqrt{2}}\end{pmatrix}=-\frac{1}{2}\begin{pmatrix}0\\\frac{v}{\sqrt{2}}\end{pmatrix}\neq 0\\
    &Y\langle \Phi\rangle=\frac{1}{2}\langle\Phi\rangle\neq 0.
        \end{split}
\end{equation}
All the generators are broken, but there is a surviving symmetry associated with the electric charge:
\begin{equation}
    Q\langle \Phi\rangle=(Y+T_3)\langle\Phi\rangle=\left(\frac{1}{2}-\frac{1}{2}\right)\langle\Phi\rangle=0.
\end{equation}
As we can see, the
vev is neutral and the charge symmetry is conserved. Therefore, the gauge symmetry is spontaneously broken in the following way:
\begin{equation}
    SU(3)_c\times SU(2)_L\times U(1)_Y\to SU(3)_c\times  U(1)_{Q}.
\end{equation}
Taking into account the symmetry breaking, the Higgs doublet can be rewritten using the Kibble transformation:
\begin{equation}
    \Phi=\frac{\exp\left(i\sum_aT_a'\epsilon^a(x)\right)}{\sqrt{2}}\begin{pmatrix}0\\H(x)+v\end{pmatrix},
\end{equation}
where $T_a'$ are the broken generators ($T_1,T_2,T_3-Y$) and the $\epsilon^a$ are the associated Goldstone bosons. The only physical degree of freedom is $H$, which is kwown as the Higgs boson. With all these considerations, the lagrangian for the Higgs sector can be written as follows:
\begin{equation}
    \mathcal{L}=(D_\mu \Phi)^\dagger D^\mu \Phi-V(\Phi).
\end{equation}
\subsection{The Yukawa interaction}
The Higgs doublet can couple to fermions via the Yukawa coupling. To show the basic structure of this interaction, we consider the first family of leptons:
\begin{equation}
    \mathcal{L}_{yuk}=-y_e\bar{L}\Phi e_R-y_\nu \bar{L}\tilde{\Phi} \nu_R +h.c.
\end{equation}
With $\tilde{\Phi}=i\sigma_2\Phi^*$. The non vanishing vev generates the masses of the fermions, which are proportional to the Higgs couplings to fermions. In our framework, we considered only the first family, but if we extend to the rest of families, we get the possibility of mixings:

\begin{equation}
    \mathcal{L}_{yuk}=-\sum_{i,j}^{families}y_{ij}^e\bar{L}^i\Phi e_R^j+y_{ij}^\nu \bar{L}^i\tilde{\Phi} \nu_R^j +\text{h.c.}
\end{equation}
After the spontaneous symmetry breaking, the mass matrix for a given fermion $f$ has the following form:
\begin{equation}
    m_{ij}^{f}=\frac{y_{ij}^f v}{\sqrt{2}}.
\end{equation}
As we can see, the SM predicts a direct relation between fermion masses and the corresponding couplings with the Higgs.

\section{Electroweak symmetry breaking and the physical fields}
In this section, we show the effect of SSB in the gauge fields. The important terms come from the kinetic term of the Higgs doublet:
\begin{equation}
    (D_{\mu} \Phi)^\dagger D^{\mu}\Phi=\frac{1}{2}\begin{pmatrix}0&v\end{pmatrix}\left(\frac{g}{2}\sigma_a W_{\mu}^a+\frac{g'}{2}B_\mu\right)^2\begin{pmatrix}0\\v\end{pmatrix}+\text{Higgs-gauge couplings}.
\end{equation}
We will focus on the vev term, and expand as follows:
\begin{equation}
\begin{split}
  [(D_{\mu} \Phi)^\dagger D^{\mu}\Phi]^{(vev)}&=\frac{1}{8}\begin{pmatrix}0&v\end{pmatrix}\begin{pmatrix}gW_\mu^3+g'B_\mu&g(W_\mu^1-iW_\mu^2)\\g(W_\mu^1+iW_\mu^2)&-gW_\mu^3+g'B_\mu\end{pmatrix}^2\begin{pmatrix}0\\v\end{pmatrix}\\
   &=\frac{g^2v^2}{8}(W_\mu^1-iW_\mu^2)(W^{1\mu}+iW^{2\mu})+\frac{v^2}{8}\begin{pmatrix}W_\mu^3&B_\mu\end{pmatrix}\begin{pmatrix}
   g^2&-gg'\\-gg'&g'^2
   \end{pmatrix}\begin{pmatrix}W^{3\mu}\\B^\mu\end{pmatrix}.
   \end{split}
\end{equation}
We can see that the vev generates mass terms for the gauge bosons. The $W^1$ and $W^2$ behave like real and imaginary parts of a complex field, therefore we define:
\begin{equation}
    W_\mu^{\pm}=\frac{W_\mu^1\mp iW_\mu^2}{\sqrt{2}},
\end{equation}
while $W^3$ and $B$ mix by a non diagonal mass matrix. It's useful to define the physical fields as the eigenstates of the mass matrix. Doing so we find that:
\begin{equation}
    \begin{pmatrix}Z_\mu\\A_\mu\end{pmatrix}=\begin{pmatrix}\cos\theta_W&-\sin\theta_W\\
    \sin\theta_W &\cos\theta_ W\end{pmatrix}\begin{pmatrix}W_\mu^{3}\\B_\mu\end{pmatrix},
\end{equation}
where we define the weak mixing angle as:
\begin{equation}
    \tan\theta_W=\frac{g'}{g}.
\end{equation}
The mass spectrum can be rewritten in terms of the physical fields:
\begin{equation}
\begin{split}
    [(D_{\mu} \Phi)^\dagger D^{\mu}\Phi]^{(vev)}&=\frac{g^2v^2}{4}W_\mu^+W^{-\mu}+\frac{v^2}{8}\begin{pmatrix}Z_\mu &A_\mu\end{pmatrix}\begin{pmatrix}g^2+g'^2&0\\0&0\end{pmatrix}\begin{pmatrix}Z^\mu \\A^\mu\end{pmatrix}\\
    \implies &M_W=\frac{gv}{2}\\
    &M_Z=\frac{(\sqrt{g^2+g'^2})v}{2}\\
    &M_A=0.
    \end{split}
\end{equation}
The $W^\pm$ and $Z$ bosons mediate the weak interactions, whereas $A$ corresponds to the photon that mediates electromagnetism. It's worth to note the direct relation between massive vector bosons and broken symmetries, while, the strong force carriers (also known as gluons due to historical reasons) and photons remain massless, the weak force carriers acquire mass.

\section{Success of the Standard Model}
\begin{figure}[!h]
    \centering
    \includegraphics[width=0.6\textwidth]{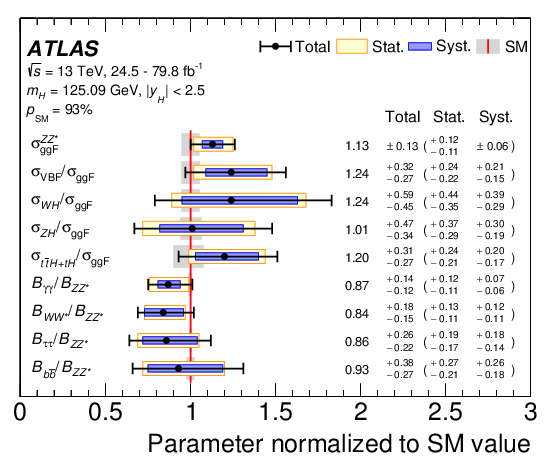}
    \caption{Comparison between Higgs boson measurements and the corresponding SM prediction. Taken from Ref. \cite{higgsmeas}.}
    \label{success}
\end{figure}

As we stated before, the standard model is a theory of interactions. All the couplings constants are free parameters, and their values can not be extracted from the theory. However, after years of experiments and measurements the particle physics community has been able to apply global fits on these parameters, showing that the model is consistent. As an example of the model success, we consider the results from Ref. \cite{higgsmeas}. In this reference, several measurements on the Higgs boson production and decay are performed, all of them in agreement with the SM predictions, as can be seen in Figure \ref{success}. This result is one of many confirmations of the SM. A general review containing all the measurements of particle physics observables can be found in Ref. \cite{pdg2022}. 

Undoubtedly, this theory has been remarkably successful, but there are some observed phenomena that cannot be explained in the framework of the SM. These observed phenomena are studied in the next chapter.

\chapter{Some problems of the Standard Model}\label{smproblems}
Despite the indisputable success of the SM as a description of fundamental interactions, there are some physical phenomena that this theory can not explain. 
These are, among others, the nature of dark matter (DM) and the neutrino mass generation mechanism. In this chapter, we review these two problems, focusing on the evidence and the reasons why we should study theories beyond the SM.
\begin{section}{Dark Matter}
In this section, we review the key concepts related to the Dark Matter problem. In Section \ref{astroevidence} we review the evidence coming from astrophysical observations on galaxies, while we review the cosmological evidence in section \ref{cosmoevidence}. 
The entire section is mainly based on Ref. \cite{majumdar}.
\subsection{Astrophysical evidence of DM}\label{astroevidence}
The DM presence is inferred from the study of celestial objects such as galaxies. Some tensions between the current understanding of celestial dynamics and the direct observations of these structures have motivated the hypothesis of a non luminous type of matter that affects the dynamics of stars from \emph{the dark}. Among these tensions, we can mention the following: The flat rotation curves, the structure of galaxy clusters and the gravitational lensing effect on images.  
\subsubsection{Rotation Curves of Spiral Galaxies} \label{rotcurves}
The spiral galaxies can be well described by means of non relativistic Classical Mechanics, due to the large distance between the galactic center and the stars forming the spiral. In this sense, the stars rotate in circular motion around the galactic center, with a centripetal force related to the gravitational interaction:
\begin{equation}
    \frac{mv^2(r)}{r}=\frac{Gm}{r^2} \int_{0}^r\rho(r')\mathrm{d}r',
\end{equation}
where $\rho(r)$ is the mass density of the galaxy. From this equation it's possible to find the radial dependence on the orbital velocity of the stars. This dependence is called Rotation Curve of a galaxy.
From astronomical observations, the mass density is expected to decrease with the distance, but the measurements of the rotation curves show a flat behaviour for large distances, as can be seen in Figure \ref{flat_rot}. The explanation to this anomaly is the presence of non visible matter surrounding the galaxy, forming a dark halo.
\begin{figure}[!h]
    \centering
    \includegraphics[width=0.5\textwidth]{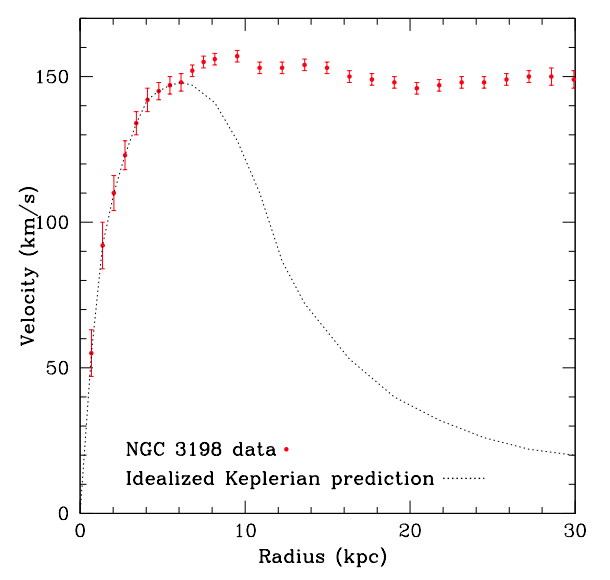}
    \caption{Galaxy rotation curve for a spiral galaxy, taken from Ref. \cite{primer}.}
    \label{flat_rot}
\end{figure}
\subsubsection{Galaxy Clusters} 
A galaxy cluster is a bound state formed by a group of galaxies. A system of this class can be described by the virial theorem:
\begin{equation}
    V+2T=0,
\end{equation}
where $V$ is the potential energy of the system and $T$ is the kinetic energy of the system. A galaxy cluster can be described as a sphere of mass $M$ and radius $R$, with a uniform density $\rho$. The potential energy due to self-gravitation can be easily calculated:
\begin{equation}
    V=-G\int\frac{1}{r}\left(\frac{4}{3}\pi r^3\rho\right)\left(4\pi r^2\rho\right)\mathrm{d}r=-\frac{3}{5}\frac{GM^2}{R},
\end{equation}
while the kinetic energy of the system can be estimated as:
\begin{equation}
    T=\frac{1}{2}Mv_{rms}^2,
\end{equation}
where $v_{rms}$ stands for the the root mean square speed of the galaxies forming the cluster. With this expressions it's possible to find a relationship between $M$, $R$ and $v_{rms}$:
\begin{equation}
  M=\frac{5}{3}\frac{R}{G}v_{rms}^2.
\end{equation}
From the measurements of $R$ and $v_{rms}$ it's possible to estimate the gravitating mass, which is confronted with the luminous mass, giving a discrepancy that can only be explained with the presence of DM. According to Ref. \cite{majumdar}, the study of the Coma Cluster performed by Zwicky in 1933 showed the estimation of the gravitating mass, and defined the mass-to-luminosity ratio $M/L$. Zwicky computed this ratio for the cluster and for each galaxy forming the cluster, finding the following relation:
\begin{equation}
    \frac{(M/L)_{cluster}}{(M/L)_{galaxy}}\sim 50.
\end{equation}
The difference can be explained by the presence of DM filling the space between galaxies, supporting the explanation of the DM halo presented in section \ref{rotcurves}.
\subsubsection{Gravitational Lensing}
Up to now, all the DM evidence has been justified by the use of Newtonian Mechanics, but there is a phenomenon called Gravitational Lensing, which corresponds to the bending of light trajectory due to the presence of gravitating mass. This phenomenon is described by means of General Relativity. Since the bending of light is related to the mass, this effect can be used to describe in a more accurate way the distribution of DM, and in some way 'seeing' it. This phenomenon was used to study the cluster 1E0657-56, also known as the bullet cluster. The studies of this cluster are the strongest evidence of DM in the context of astrophysics. It's worth to mention that dedicated studies on gravitational lensing have helped to make an actual map for the mass distribution in the Universe (Ref. \cite{DESweaklensing})
\subsection{Cosmological evidence: Relics of the Big Bang}\label{cosmoevidence}
\subsubsection{Cosmic Microwave Background fluctuations}
One of the problems of modern Cosmology is the study of the anisotropies of the Cosmic Microwave Background (CMB). These anisotropies have been measured by WMAP (Wilkinson Microwave Anisotropy Probe), obtaining the following values for the baryonic density and the total matter density of the universe, which are, respectively: $\Omega_bh^2 = 0.02260 \pm 0.00053$ and 
$\Omega_mh^2=0.1334_{−0.0055}^{+0.0056}$ (Ref. \cite{primer}). 
The best explanation until now about this discrepancy is the presence of a type of matter that doesn't interact with photons and therefore can be responsible for structure formation before the photon-baryon decoupling. As can be seen, DM appears as a natural solution to a problem in Cosmology. \newline
In order to probe the DM hypothesis, the Weakly Interacting Massive Particle (WIMP) paradigm is an interesting solution from the Particle Physics point of view. 
\subsubsection{Thermal DM and WIMPs}
If DM is a particle that was in thermal and chemical equilibrium, the production and annihilation process occur at equal rates. The process of creation and destruction is performed until the expansion of the Universe surpasses the interaction rate. The critical point when DM decouples is called freeze-out.  After the freeze-out, the DM density becomes a constant. This behaviour can be well described by the solution of the Boltzmann equation:
\begin{equation}\label{boltzmann}
    \frac{\mathrm{d}n_\chi}{\mathrm{d}t}=-3Hn_\chi-\langle\sigma v\rangle(n_\chi^2-(n_\chi)_{eq}^2),
\end{equation}
where $n_\chi$ is the DM number density, $H$ is the Hubble constant and $\langle\sigma v\rangle$ is the thermally average product between the annihilation cross section and the relative velocity between the DM particles. This quantity is the link between Cosmology and Particle Physics, because it determines the DM density after the freeze-out. After solving the Boltzmann equation, the DM density can be estimated as:
\begin{equation}\label{reliceq}
    \Omega_\chi h^2\approx \frac{3\times10^{-27}[cm^3 s^{-1}]}{\langle\sigma v\rangle}.
\end{equation}
While the measured value for this quantity is $\Omega_\chi h^2=0.11425\pm 0.00311$ (Ref. \cite{primer}).
The effect of the annihilation cross section can be seen in Figure \ref{relic}. From the Particle Physics perspective, eq. \eqref{reliceq} allows to falsify the WIMP hypothesis for a given DM particle candidate. 

\begin{figure}[!h]
    \centering
    \includegraphics[width=0.45\textwidth]{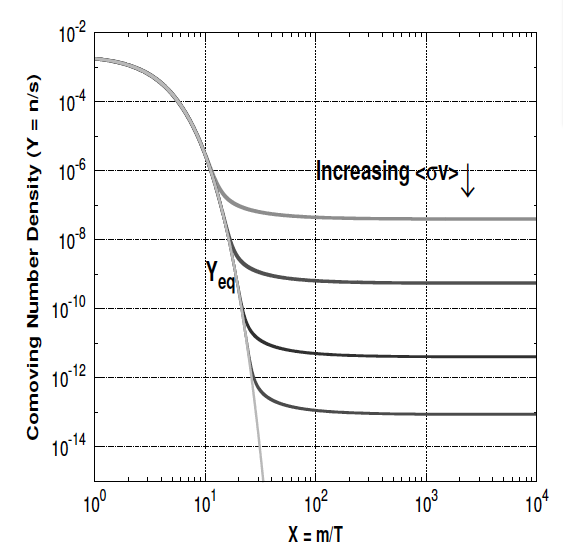}
    \caption{Evolution of the DM density as a function of $x=m/T$. The number density is normalized by the entropy density of the universe, $s$. Taken from Ref. \cite{majumdar}.}
    \label{relic}
\end{figure}

\subsubsection{The need of going beyond the Standard Model}
After checking out the evidence from Ref. \cite{majumdar}, it is clear that a significant part of the mass density in the universe comes from a non luminous type of matter. Could dark matter be contained in the SM? A natural guess would be that neutrinos, the only neutral fermions in the SM could be the answer to the dark matter problem. However, these particles have a very tiny mass, which make them relativistic particles. The current observations suggest that dark matter is \emph{cold}, in the sense that it's motion is non relativistic. Therefore, relativistic particles can not explain the DM abundance under the current cosmological paradigm. Even if we ignore the fact that neutrinos are relativistic particles, the neutrino relic density is way lower than the DM relic density. According to Ref. \cite{primer}, the neutrino relic density is bounded by $\Omega_\nu h^2< 0.0072$. Hence, the abundance of dark matter in the universe cannot be explained by the Standard Model.

\end{section}

\begin{section}{Neutrino Physics}
Up to now, most of the predicted interactions in the SM have been proven, except of the interaction between neutrinos and the Higgs boson. Since the mass of these particles is very small ($<1.1$ [eV] according to Ref. \cite{pdg2022}), the Higgs interaction is highly suppressed. On the other hand, the interaction between the Higgs and neutrinos needs the presence of left and right handed neutrinos, but the latter ones haven't been detected. The detection of right handed neutrinos is particularly hard because this particle is a SM singlet. The only way of detecting them would be gravitational interactions. The lack of evidence on  right handed neutrinos motivates the study of alternative neutrino mass generation mechanisms beyond the SM. In order to study the different descriptions of neutrino mass terms, we are following Ref. \cite{neutrinos}.
\subsection{Different types of mass terms}
In order to describe the different mass terms that can be built, we consider a set of general fermions $\{\psi_l\}$, with $l\in \{1,2,\dots, n\}$. If we allow these fermions to mix, the mass terms can be written as a  composed by a $n\times n$ matrix.
\subsubsection{Dirac mass}
In the SM all fermions have a Dirac mass term:
\begin{equation}
    \mathcal{L}_{D}=-\sum_{ll'}\bar{\psi}_R^lM^D_{ll'}\psi_L^{l'}+\text{h.c.}
\end{equation}
As can be seen, a Dirac mass term requires the existence of both left and right components of the fermion.

\subsubsection{Majorana mass}
Another possible way of defining a mass term is by the so called Majorana mass term. 
It's useful to define the conjugation of a field as:
\begin{equation}
\begin{split}
    &\psi_L^c=C\bar{\psi}_L^T \\
    &\psi_R^c=C\bar{\psi}_R^T. 
    \end{split}
\end{equation}
For one side, we have that $\gamma_5\psi_L=-\psi_L$ by construction.
It's easy to see that the $\psi_L^c$ is the right handed partner of $\psi_L$. The demostration is straightforward:
\begin{equation}
\begin{split}
    \gamma_5\psi_L^c&= \gamma_5C\bar{\psi}_L^T \\
    &=C\gamma_5^T\bar{\psi}_L^T\\
    &=C(\bar{\psi}_L\gamma_5)^T\\
    &=C(\bar{\psi}_L)^T=\psi_L^c\\
    &\therefore \gamma_5\psi_L^c=\psi_L^c.
    \end{split}
\end{equation}
Here we used basic Dirac algebra. Since the field conjugation changes the field helicity, we can build a mass term without making reference to an explicitly right handed component.  The mass term built with this field conjugation is called a Majorana term. The following example shows the structure of a typical Majorana mass term:
\begin{equation}
    \mathcal{L}_M=-\frac{1}{2}\sum_{ll'}\bar{\psi}_L^lM^M_{ll'}\psi_L^{l'c}+\text{h.c.}
\end{equation}
The same mass term can be built for right handed fermions, keeping the same structure. 
\subsection{The seesaw mechanism}
Both Dirac and Majorana terms can coexist. To show the effect of both Dirac and Majorana masses, we consider only one family of fermions with two fermion fields (left and right):
\begin{equation}\label{eq:Ld+m}
    \mathcal{L}^{D+M}=-\frac{1}{2}m_L\bar{\psi}_L\psi_L^c-m_D\bar{\psi}_L\psi_R-\frac{1}{2}m_R\bar{\psi^c}_R\psi_R+\text{h.c.}
\end{equation}
The lagrangian can be rewritten considering the following definitions:
\begin{equation}
    \Psi_L=\begin{pmatrix}\psi_L\\\psi_R^c\end{pmatrix}\quad \text{and}\quad M^{D+M}=\begin{pmatrix}m_L&m_D\\m_D&m_R\end{pmatrix}.
\end{equation}
With these definitions, the lagrangian of eq. \eqref{eq:Ld+m} becomes:
\begin{equation}
    \mathcal{L}^{D+M}=-\frac{1}{2}\bar{\Psi}_LM^{D+M}\Psi_L^c+\text{h.c.}
\end{equation}
Clearly, the $\Psi$ field is a Majorana field and exists a mixing. The physical fields propagate with definite masses which are eigenvalues of $M^{D+M}$. 
These eigenvalues are:
\begin{equation}\label{eigenmasses}
    m_{1,2}=\frac{1}{2}(m_L+m_R)\mp\frac{1}{2}\sqrt{(m_R-m_L)^2+4m_D^2}.
\end{equation}
This mixing motivates the introduction of the seesaw mechanism in the SM to explain the neutrino mass generation mechanism. In the SM the introduction of a left-handed Majorana term would violate gauge charges, so we need to impose $m_L=0$ to respect unbroken symmetries. Since right-handed neutrinos are gauge singlets, there is no problem on introducing a Majorana mass term, while the Dirac component is supposed to come from a standard Yukawa interaction. The key assumption in the Seesaw mechanism is that $m_R>> m_D$. Expanding eq. \eqref{eigenmasses} in terms of $m_D^2/m_R^2$ we get:
\begin{equation}
    m_{1,2}\approx \frac{1}{2}m_R\mp\frac{m_R}{2}\left(1+2\frac{m_D^2}{m_R^2}\right)\implies |m_1|\approx m_R \quad \text{and}\quad |m_2|=\frac{m_d^2}{m_R}.
\end{equation}
The relation between the physical states and the original ones are:
\begin{equation}
    \begin{split}
        &\psi_L=i\psi_{1L}+\frac{m_D}{m_R}\psi_{2L}\\
        &\psi_R^c=-i\frac{m_D}{m_R}\psi_{1L}+\psi_{2L}.
    \end{split}
\end{equation}
Therefore, the active neutrinos acquire naturally a small mass suppressed by $m_R$, which can be arbitrarily large. The seesaw mechanism explains the small neutrinos masses in an economical way, but it depends entirely on $m_R$, which is expected to be significantly large. On the other side, the right handed sterile neutrino is still a singlet under the SM, which means it has no gauge interactions. The lack of gauge interactions implies that the sterile neutrino hypothesis has the same problem of falsifiability as the Dirac neutrino hypothesis.
\subsection{Radiative neutrino masses}
An interesting possibility for explaining the tiny neutrino masses relies in considering that it doesn't exist a mass term for neutrinos, and their mass come from radiative processes. The radiative mechanisms vary depending on the nature of neutrino masses. A detailed explanation of these different mechanisms can be found in Ref. \cite{radiative_mass}. We will focus on Majorana radiative mass models because they don't need a right handed neutrino, bypassing the problem of right handed neutrino detection. 
\subsubsection{The Weinberg operator}
If neutrinos are Majorana particles, they can acquire mass via the Weinberg operator, which has the folloding form (Ref. \cite{langacker}):
\begin{equation}
    \mathcal{O}_W=\frac{1}{\Lambda}(L\overrightarrow{\tau}\tilde{L}_R^c)(\phi^\dagger \overrightarrow{\tau}\tilde{\phi})+ \text{h.c.}, \quad \text{with}\quad \tilde{L}_R^c=\begin{pmatrix}e_R^c\\-\nu_R^c\end{pmatrix}.
\end{equation}
This is a dimension 5 operator which breaks lepton number conservation. Therefore, it's expected to be an effective operator suppressed by the scale $\Lambda$ where lepton number symmetry breaks. This term can be obtained by integrating out new physics appearing at higher scales.
At the practical level, a radiative neutrino mass term must raise from loop calculations where the particles running in the loop can be SM and/or new physics. One possibility is to consider SM extensions that can solve more than one SM problem.
\subsubsection{The scotogenic scheme}
As we stated before, neutrino masses can be generated by effective operators related with new physics. There is a type of models where the new particles generating neutrino masses can be DM candidates. We refer to this type of models as  \emph{scotogenic} models for neutrino mass. The scotogenic hypothesis can be falsified by applying constraints on the parameter space from both DM studies and neutrino experiments. If the hypothesis is correct, the parameter space constraints should converge. Therefore, from the phenomenological point of view,  theories with scotogenic neutrino masses are interesting and have the chance of being probed or disregarded in the near future.

\end{section}

\chapter{The proposed model}
The need of physics beyond the standard model has motivated the rise of a large number of models, with different levels of complexity. Among them, it's worth to highlight the minimal dark matter program (Ref. \cite{minimaldm}). In this context, there is a family of models with a common paradigm: the addition of an electroweak multiplet. Depending on the quantum number assignation, these multiplets may contain one or more DM candidates. This multiplets can be scalars, fermions or vectors. 
In this chapter, we study an extension of the SM composed by a massive vector $SU(2)_L$ doublet and a Left-handed Heavy Neutral Lepton (HNL for short). We review the main structure of the model and some results that attempt to solve the problems of the SM that were presented in the previous chapter.
\section{The anatomy of the Model}
The model history has two main parts: originally, the authors of  Ref. \cite{vector_dm} introduced the vector sector in the context of minimal dark matter models, without considering the left handed HNL. However, the addition of this exotic fermion is a natural extension due to the vector doublet quantum numbers. Therefore, the model description is separated by the chronological order in which the different sectors were presented in the literature.
\subsection{The Vector Sector}
The Vector sector of the model was first proposed in Ref. \cite{vector_dm}. The main ingredient is the addition of a Vector Doublet $V_\mu$:
\begin{equation}
    V_{\mu}=\begin{pmatrix}V^{+}_\mu\\V^0_\mu\end{pmatrix}=\begin{pmatrix}V^{+}_\mu\\\frac{1}{\sqrt{2}}(V^1_\mu+iV^{2}_\mu)\end{pmatrix}.
\end{equation}
This vector doublet transforms as $(1,2,1/2)$ under the SM group $SU(3)_c\times SU(2)_L\times U(1)_Y$. The most general lagrangian that preserves gauge invariance and renormalizability has the following form:
\begin{equation}
\begin{split}
    \mathcal{L}_{DM}=&-\frac{1}{2}(D_\mu V_\nu -D_\nu V_\mu)^\dagger (D^\mu V^\nu- D^\nu V^\mu)+M_V^2 V_\mu^\dagger V^\mu
    \\&-\frac{1}{\xi}(D_\mu V^\mu)^\dagger(D_\nu V^\nu)+i\frac{g'\kappa_1}{2}V_\mu^\dagger B^{\mu\nu}V_\nu+ig\kappa_2 V_\mu^\dagger W^{\mu\nu}V_\nu \\
    &-\alpha_2(V_\mu^\dagger V^\mu)(V_\nu^\dagger V^\nu)-\alpha_3(V_\mu^\dagger V^\nu)(V_\nu^\dagger V^\mu)-\alpha_1[\Phi^\dagger(D_\mu V^\mu)+(D_\mu V^\mu)^\dagger \Phi]\\&-\lambda_2(\Phi^\dagger \Phi)(V_\mu^\dagger V^\mu)-\lambda_3(\Phi^\dagger V_\mu)(V^{\mu\dagger}\Phi)-\frac{\lambda_4}{2}[(\Phi^\dagger V_\mu)(\Phi^\dagger V^\mu)+(V^{\mu\dagger}\Phi)(V_\mu^\dagger \Phi)],
    \end{split}
\end{equation}
where $D_\mu$ stands for the covariant derivative, $W^{\mu\nu}$ and $B^{\mu\nu}$ are the field strengths of $SU(2)_L$ and $U(1)_Y$, respectively, and $\Phi$ is the SM Higgs doublet. This lagrangian was labeled by the authors as the Vector Doublet Dark Matter Model (VDDMM). It's worth to highlight some properties of this lagrangian.
\subsubsection{Non minimal gauge interactions} 
The non minimal gauge couplings come from the terms proportional to $\kappa_1,\kappa_2$ and $1/\xi$. The authors of Ref. \cite{vector_dm} set $\kappa_1=\kappa_2=1$ in order to avoid interactions between the photons and the neutral components of the vector field. Nevertheless, this restriction can be full filled with a more general condition: $\kappa_1=\kappa_2=\kappa$. Since the phenomenology of the model has been studied with $\kappa=1$, we are going to keep this choice, despite the fact that a different choice of this value can produce an interesting phenomenology.
\newline 
On the other side, the original model didn't contain the $1/\xi$ term, which was firstly included in Ref. \cite{vietnam}. This term allows to avoid divergences in radiative processes (see Section \ref{neumass}), but it also modifies the interaction between electroweak bosons and the vector doublet. In order to keep compatibility with the phenomenological results of Ref. \cite{vector_dm}, we are going to consider the limiting case where $1/\xi\to 0$.

\subsubsection{Accidental Symmetry}
Another interesting property of this lagrangian is the emergence of a $Z_2$ symmetry if $\alpha_1=0$. This accidental symmetry allows the lightest neutral component of the vector doublet to be a DM candidate. Taking into account these considerations about the model parameters, the lagrangian has the following form:

\begin{equation}
\begin{split}
    \mathcal{L}_{DM}=&-\frac{1}{2}(D_\mu V_\nu -D_\nu V_\mu)^\dagger (D^\mu V^\nu- D^\nu V^\mu)+M_V^2 V_\mu^\dagger V^\mu
    \\&+\kappa\big[i\frac{g'}{2}V_\mu^\dagger B^{\mu\nu}V_\nu+ig V_\mu^\dagger W^{\mu\nu}V_\nu\big] -\alpha_2(V_\mu^\dagger V^\mu)(V_\nu^\dagger V^\nu)-\alpha_3(V_\mu^\dagger V^\nu)(V_\nu^\dagger V^\mu)\\&-\lambda_2(\Phi^\dagger \Phi)(V_\mu^\dagger V^\mu)-\lambda_3(\Phi^\dagger V_\mu)(V^{\mu\dagger}\Phi)-\frac{\lambda_4}{2}[(\Phi^\dagger V_\mu)(\Phi^\dagger V^\mu)+(V^{\mu\dagger}\Phi)(V_\mu^\dagger \Phi)].
    \end{split}
\end{equation}

\subsubsection{Parameter definition}
In order to perform calculations, it's useful to define some quantities after the symmetry breaking:
\begin{equation}
    \begin{split}
        &M_{V^\pm}^2=\frac{1}{2}(2M_V^2-v^2\lambda_2)\\
        &M_{V^1}^2=\frac{1}{2}(2M_V^2-v^2[\lambda_2+\lambda_3+\lambda_4])\\
        &M_{V^2}^2=\frac{1}{2}(2M_V^2-v^2[\lambda_2+\lambda_3-\lambda_4])\\
        &\lambda_L=\lambda_2+\lambda_3+\lambda_4,
    \end{split}
\end{equation}
where $\lambda_L$ stands for the physical coupling between $V^1$ and the Higgs Boson. For practical reasons, it's possible to rewrite the lagrangian in terms of these parameters. This can be done by inverting the set of defining equations. The authors of Ref. \cite{vector_dm} already performed this calculation, obtaining: 
\begin{equation}
\begin{split}
    &\lambda_2=\lambda_L+2\frac{M_{V^1}^2-M_{V^\pm}^2}{v^2}\\
    &\lambda_3=\frac{2M_{V^\pm}^2-M_{V^1}^2-M_{V^2}^2}{v^2}\\
    &\lambda_4=\frac{M_{V^2}^2-M_{V^1}^2}{v^2}\\
    &M_V^2=M_{V^1}^2+\frac{v^2}{2}\lambda_L.
    \end{split}
\end{equation}
Due to the vector doublet quantum numbers and Lorentz invariance, it is not possible to link directly the new vector field to the SM fermions. However, these particles can be linked if we introduce an exotic left-handed neutrino which is singlet of the SM,
 which is presented in the next section.
\subsection{The Left-Handed Singlet Fermion Sector}
The minimal lagrangian for a new fermion state that can link the vector doublet and SM leptons has the following form:
\begin{equation}
    \mathcal{L}_{N_L}=\frac{1}{2}(i\bar{N}_L^c\gamma^\mu \partial_\mu N_L -M_N\bar{N}_L^cN_L)-\sum_{k=\{e,\mu,\tau\}}\beta_k \bar{L}_k \gamma^\mu\tilde{V_\mu} N_L  +\text{h.c.},
\end{equation}
with
\begin{equation}
    \tilde{V_\mu}=i\sigma_2 V_\mu^*=\begin{pmatrix}\frac{V^1_\mu-iV^{2}_\mu}{\sqrt{2}}\\-V^{-}_\mu\end{pmatrix}.
\end{equation}
In order to respect gauge symmetry, the new fermion $N_L$ must transform as $(1,1,0)$ under the SM group. Since the new fermion is a singlet, we choosed it to be a Majorana particle,  doing so we avoid the introduction of a right-handed partner, keeping the model as minimal as possible. This exotic fermion can be labeled as a left-handed Heavy Neutral Lepton (HNL). Additionally, we choose $N_L$ to be odd under the emergent $Z_2$ symmetry from the vector sector. \newline
Extensions to the VDDMM with fermion states of this kind have been studied with two (Ref. \cite{masses_and_mixings}) and three (Ref. \cite{vietnam}) left handed HNLs. However, to keep a minimalistic philosophy we are going to consider the addition of only one HNL. In order to keep track on the different models, we denoted the complete model as VDM1N.
\section{The model capabilites}
The addition of interactions between the new particles and SM fermions allows to find possible solutions to the problems presented in the previous chapter. In this section, we review how this model can fit solutions to these problems.
\subsection{Dark Matter}
The model has two DM candidates, depending on the mass hierarchy between the vector doublet neutral components and the HNL. If we consider the HNL as the DM candidate, the annihilation cross section has contributions only from t-channel interactions with the SM leptons. This case was studied in Ref. \cite{vietnam}, obtaining:
\begin{equation}\label{annicross}
    \langle \sigma v\rangle=\sum_{k,k'=\{e,\mu,\tau\}}|\beta_k^*\beta_{k'}|^2\frac{M_N^2}{8\pi}\left(1+\frac{8T_f}{M_N}\right)\left(\frac{1}{M_{V^+}^4}+\frac{4}{(M_{V^1}^2+M_{V^2}^2)^2}\right),
\end{equation}
while for the other case, the annihilation cross section will have contributions from t-channel exchange of HNLs and electroweak contributions. The latter ones were studied in Ref. \cite{vector_dm}. The addition of the HNL should modify the results presented in this reference, but the study of these effects are beyond the scope of this work.
\subsection{Neutrino mass}\label{neumass}
\begin{figure}[!h]
    \centering
    \includegraphics[scale=0.4]{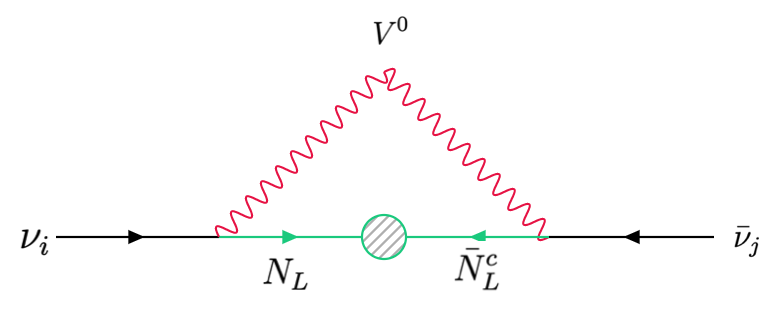}
    \caption{Diagram for radiative neutrino masses. The central circle represents a mass injection.}
    \label{neutrinomass}
\end{figure}

It's possible to produce the SM neutrino masses via loop diagrams involving the neutral components of the vector doublet and the HNL, as can be seen in Figure \ref{neutrinomass}. 
The present model has a divergence on the calculation of the loop integral, which has the following form according to Ref. \cite{masses_and_mixings}:
\begin{equation}
    M_{ij}\propto f_\Lambda(M_N,M_{V^1})-f_\Lambda(M_N,M_{V^2}),
\end{equation}
With
\begin{equation}
    f_\Lambda(x,y)=\frac{\Lambda^2}{y^2}+\frac{y^2}{y^2-x^2}+\frac{x^2}{y^2(y^2-x^2)}\ln\left(1+\frac{\Lambda^2}{x^2}\right).
\end{equation}
Here, $\Lambda$ is a cutoff due to a divergence arising from the $k^\mu k^\nu/M_{V}^2$ term of the vector propagator. This divergence can be bypassed by including a term proportional to $1/\xi$, which we neglected in the model definition. Taking into account this term, the authors of Ref. \cite{vietnam} obtain:
\begin{equation}
    M_{ij}=\frac{\beta_i\beta_j M_N}{32\pi^2}(f_\xi(M_N^2,M_{V^1}^2)-f_\xi(M_N^2,M_{V^2}^2)),
\end{equation}
with
\begin{equation}
    f_\xi(x,y)=\frac{y}{x-y} \left(\frac{((3-\xi) (1+\xi) x-\xi (3+\xi) y) \ln \left(\frac{y}{x}\right)}{x-\xi y}-\frac{\xi^2 \ln (\xi) (x-y)}{x-\xi y}\right).
\end{equation}
Both cases have the same behaviour, the neutrino masses are defined by the mass difference between $V^1$ and $V^2$. Also, from a practical point of view, the parameter $\xi$ can act as a regulator, in the same way as $\Lambda$. This quantity must be fitted so it doesn't increase the predictability of the model.

\subsection{Muon Anomalous magnetic moment}
\begin{figure}[!h]
    \centering
    \includegraphics[width=0.5\textwidth]{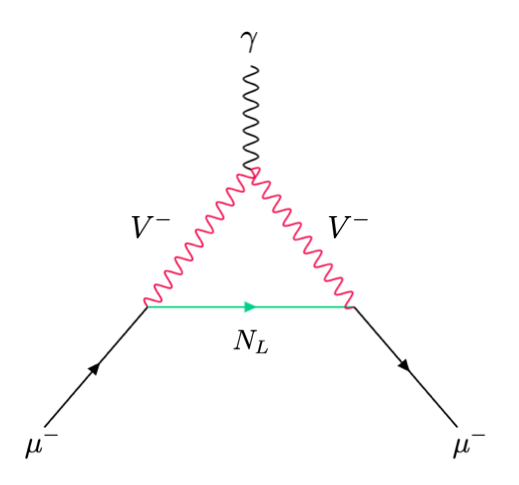}
    \caption{Model contribution to the muon anomalous magnetic moment.}
    \label{g2diagram}
\end{figure}

The discrepancy between the SM prediction for the muon anomalous magnetic moment and the measured value of $a_\mu\equiv(g-2)/2$ (Ref. \cite{muong2}) can be explained by the diagram presented in Figure \ref{g2diagram}.
The loop calculation of this diagram was performed by the authors of Ref. \cite{vietnam}, obtaining:
\begin{equation}\label{eqg2}
    \Delta a_\mu=\frac{|\beta_\mu|^2}{8\pi^2}\frac{m_\mu^2}{M_{V^+}^2}\int_0^1\frac{x(1+x)M_{V^+}^2+(1-x)(1-x/2)M_N^2}{xM_{V^+}^2+(1-x)M_N^2}x\mathrm{d}x.
\end{equation}
It's worth to mention that we didn't include this discrepancy as a problem of the SM in Chapter \ref{smproblems}. This choice is motivated by the fact that this discrepancy is below the $5\sigma$ significance threshold, which is a consensus in the Particle Physics community to claim a discovery. However, this quantity must not be ignored and the $g-2$ measurements can be used to constrain models beyond the SM, like in this case. On the other side, the authors of Ref. \cite{vietnam} don't inform the values used for the non minimal gauge couplings, so we are skeptical about the applicability of their result for parameter constraints.

\subsection{Lepton Flavor Violation}
\begin{figure}
    \centering
    \includegraphics[width=0.5\textwidth]{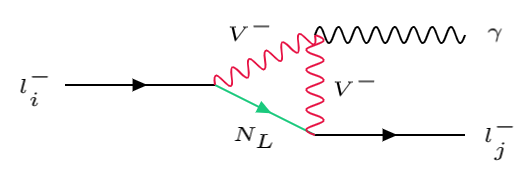}
    \caption{Diagram for LFV charged lepton decay.}
    \label{diag_lfv}
\end{figure}

The model structure allows to have lepton flavor violating (LFV) decays. For instance, the muon decay $\mu \to e \gamma$ can be produced via one loop processes. This process is highly constrained, according to Ref. \cite{pdg2022}, the LFV muon branching fraction has an upper bound $\text{Br}(\mu\to e \gamma)<4.2\times 10^{-13}$. Due to the strong experimental limits, the study of this process is a great opportunity to constrain the parameter space of the model.
The Feynman diagram for this LFV charged lepton decay can be seen in Figure \ref{diag_lfv}. It's worth to mention that this diagram has the same topology as the one depicted in Figure \ref{g2diagram} for the muon anomalous magnetic moment.

These type of processes has been studied in depth in Ref. \cite{ilakovac}, in models where heavy neutrinos mix with SM neutrinos. In this reference, the diagrams considered have $W^\pm$ bosons instead of the new vector bosons, but the topology is also the same.
However, these results can not be used directly in the context of our model due to the non minimal gauge interactions. It's worth to mention that the non minimal gauge couplings can be fixed in order to make applicable the results from Ref. \cite{ilakovac}.
\subsubsection{Charged vector magnetic moment}
LFV processes calculated in Ref. \cite{ilakovac} rely on the coupling between vector bosons and the photon. The relevant Feynman rules can be seen in Table \ref{tab_feyn}. These Feynman rules can be forced to be equal if we take the difference between them, obtaining:
\begin{equation}
    \Delta=ie [(1 + \kappa) p_ 1^\lambda g^{\alpha\beta} - (1 + \kappa) p_1^\beta g^{\alpha\lambda} +
   1/\xi (p_ 3^\lambda g^{\alpha \beta} - 
      p_ 2^\beta g^{\alpha \lambda})].
\end{equation}
Therefore, if we take $\kappa=-1$ and $1/\xi=0$, the vertices are equivalent and the results from the reference can be used. Under this framework, the expression for the branching fraction has the following form:

\begin{equation}
    \text{Br}(l_i\to l_j\gamma)_{\kappa=-1}=\frac{|\beta_i|^2|\beta_j|^2g_w^2s_w^2m_i^5}{64(4\pi)^4M_{V^+}^4\Gamma_i}|G\left(M_N^2/M_{V^+}^2\right)|^2,
\end{equation}
with 
\begin{equation}
    G(x)=-\frac{2x^3+5x^2-x}{4(1-x)^3}-\frac{3x^3}{2(1-x)^4}\ln x.
\end{equation}

However, the phenomenology of the model has been studied with $\kappa=1$ (Ref. \cite{vector_dm}), hence we are going to develop the following calculations with this consideration.
\begin{table}[!h]
    \centering
    \begin{tabular}{|c|c|}
    \hline
        Particles & Feynman rule \\
        \hline
       $A^\alpha(p_1),W^{+\beta}(p_2),W^{-\lambda}(p_3)$  & $-ie p_ 1^\lambda g^{\alpha\beta} + ie p_ 2^\lambda g^{\alpha \beta} + 
 i e p_ 1^\alpha g^{\beta\lambda} - 
 i e p_ 3^\beta g^{\alpha \lambda} - 
 i e p_ 2^\alpha g^{\beta\lambda} + iep_ 3^\alpha g^{\beta\lambda}$\\
   \hline
       $A^\alpha(p_1),V^{+\beta}(p_2),V^{-\lambda}(p_3)$  &$-ie\kappa p_ 1^\beta g^{\alpha \lambda} + 
 ie \kappa p_ 1^\lambda g^{\alpha\beta} - 
 iep_ 2^{\alpha} g^{\beta\lambda} - 
 ie( 1/\xi) p_ 2^\beta g^{\alpha\lambda} + $\\

 &$+iep_ 2^\lambda g^{\alpha\beta} + ie p_ 3^\alpha g^{\beta\lambda} - 
 ie p_ 3^\beta g^{\alpha \lambda} + ie (1/\xi) p_ 3^\lambda g^{\alpha\beta}$\\
 \hline
    \end{tabular}
    \caption{Feynman rules describing the interaction between the charged vector bosons and the photon.}
    
    \label{tab_feyn}
\end{table}

\chapter{Left-Handed HNL production at the LHC}
Up to now, we have considered some problems of the standard model and how they could be solved in the VDM1N framework. However, the model capability is not an hypothesis confirmation. In order to falsify the VDM1N hypothesis we need to study predictions that can come from this model. The Large Hadron Collider (LHC) is an experiment dedicated to the study of particle physics, and a significant part of it's program is the search for new physics beyond the standard model.
In this chapter, we present the results for the phenomenological study on the left handed HNL production at the LHC.

\section{Preliminaries and Methods}
\subsection{The production process}
\begin{figure}[!h]
    \centering
    \includegraphics[scale=0.32]{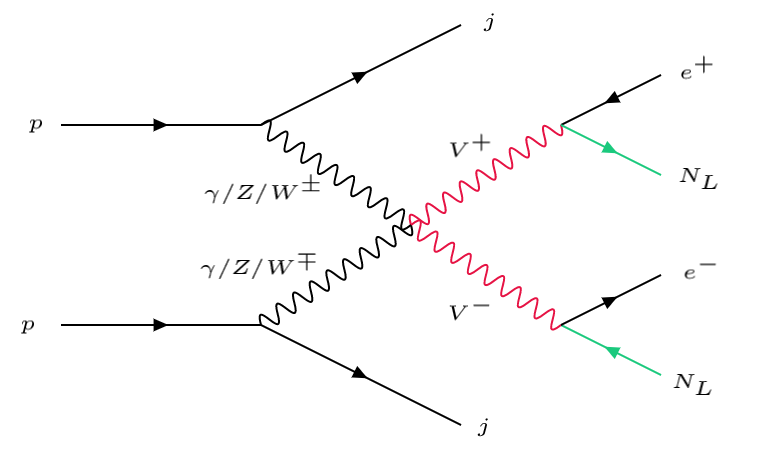}
    \caption{Leading order diagram for the production of left handed HNLs.}
    \label{vbf}
\end{figure}

Due to the HNL interactions, the only way to produce them is via vector doublet decay. Since these vectors are protected by the $Z_2$ symmetry, the double production is the only possible way. Therefore, the final state will be 2 HNLs and a lepton pair (which can be charged or neutral). The vector pair production can be done via electroweak interactions. If we consider the LHC context, the initial state is composed by two protons, so the main production mechanism will be Vector Boson Fusion (VBF) (see Figure \ref{vbf}), with the presence of at least 2 jets in the final state. The process can therefore be written as:
\begin{equation}
    p \quad p \quad\to\quad j\quad j\quad l^+\quad l^-\quad N_L\quad N_L.
\end{equation}
It's worth to mention that the HNLs could be produced accompanied by two neutrinos, but the final state would be only two jets and a significant amount of missing energy. Therefore, we prefer to study the production with charged leptons in order to have a richer signal.
We performed a computational implementation of the model. To do so, we used FeynRules (Refs. \cite{fr1,fr2,ufo}) to obtain the Feynman Rules for the new sector, and used Madgraph  (Ref. \cite{mg5}) to compute the cross section for the process. It's worth to mention the topology of the final state, which is composed by the following objects: 2 jets from the outgoing quarks, 2 Same Flavor Opposite Charge (SFOS) leptons and Missing Energy.

In principle, the final state leptons can have different flavors, but this scenario is highly constrained due to the absence of observations on Lepton Flavor Violation (LFV). This interesting possibility is not studied in this work.
\subsection{Current searches in the final state}
\subsubsection{Event Selection}
The final state of our process has been studied in Ref. \cite{susylims}. In this reference, the Signal Regions are separated in Low, Medium, and High, depending on the value of $H_T$. The common cuts for all the regions can be seen in Table \ref{cut_def}
\begin{table}[!h]
    \centering
    \begin{tabular}{|c|c|c|}
    \hline
     object    & definition &condition \\
     \hline
     $p_T(l+)$    & Transverse momentum of the positively charged lepton & $\geq 25$[GeV]\\
          $p_T(l-)$    & Transverse momentum of the negatively charged lepton & $\geq 25$[GeV]\\
          $M_{ee}$ & Invariant mass of the SFOS lepton pair & $\geq 12$[GeV]\\
          $p_T^{miss}$ & Transverse component of the missing momentum vector &$>200$[GeV]\\
          $\Delta \phi(jet_{1},p_T^{miss})$ &Azimuthal separation between $p_T^{miss}$ and the first jet & $\geq 0.4$\\
            $\Delta \phi(jet_{2},p_T^{miss})$ &Azimuthal separation between $p_T^{miss}$ and the second jet & $\geq 0.4$\\
          \hline
    \end{tabular}
    \caption{Event selection criteria. These cuts were taken from Ref. \cite{susylims}}
    \label{cut_def}
\end{table}
While there are additional cuts on SR-Medium and SR-High, with the conditions $H_T>400$[GeV] and $H_T>700$[GeV], respectively. 
In order to make our work helpful for future searches in this final state, we will apply the common cuts for all signal regions (the ATLAS cuts, from now on) over our simulated samples, as will be seen in section \ref{sec:results}

\section{Results}\label{sec:results}
\subsection{Simulated samples}
To run the simulation, we considered a special case of the process, where the SFOS pair is an electron pair. This is done because the masses of the SM leptons are much lower than the masses of the new particles, therefore the simulation shouldn't be affected by the lepton pair choice.
The simulation was used to make a scan over the values of $M_{V^+}$. We considered the special case of $M_{V^+}=M_{V^1}=M_{V^2}$. The relevant parameters were chosen as can be seen in Table \ref{paramschoice}.
\begin{table}[!h]
    \centering
    \begin{tabular}{|c|c|}
    \hline
       variable  & value  \\
       \hline
        $M_{V^+}$ & $200-300[$GeV$]$ with steps of $5[$GeV$]$\\
        $M_N$ &$50 [$GeV$]$\\
        $|\beta_e|^2$& 1\\
        $\lambda_L$& 5\\
        \hline
    \end{tabular}
    \caption{Parameters used for the simulations.}
    \label{paramschoice}
\end{table}
\newline
 In order to compare the madgraph simulations with current experimental searches, we need to define an effective cross section for the process
 \begin{equation}
     \sigma_{eff}=\mathcal{A}\epsilon \sigma,
 \end{equation}
 Where $\mathcal{A}$ stands for the acceptance, defined as the ratio of events matching the selection criteria and the total number of simulated events, and $\epsilon$ parametrizes the detector efficiency.
 \subsubsection{Detector efficiency }
\begin{figure}[!h]
     \centering
     \includegraphics[width=\textwidth]{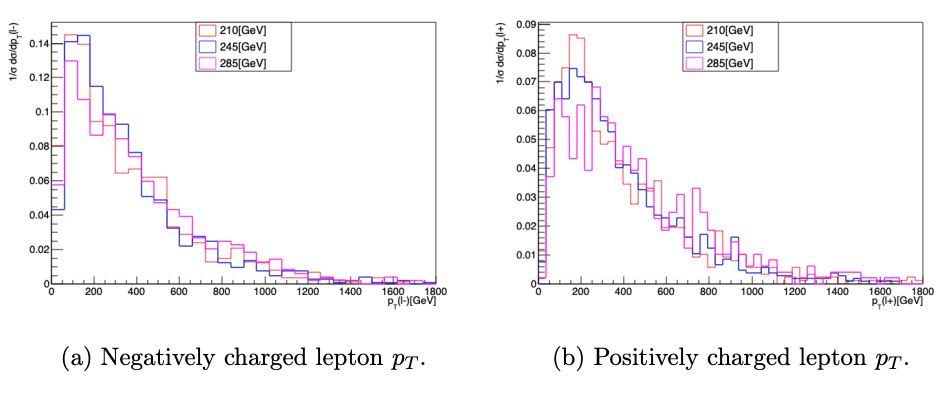}
     \caption{$p_T$ distributions for the signal leptons}
     \label{leps_pt}
 \end{figure}

 According to Ref. \cite{lep_effs}, the detector efficiency for electron reconstruction increases with the electron $p_T$ reaching values close to $1$ for values $p_T>50$[GeV]. The event selection removes all events with $p_T(l\pm)<25$[GeV], while the number of events with $25<p_T(l\pm)<50$[GeV] is very small, as can be seen in Figure \ref{leps_pt}. Therefore, we can approximate $\epsilon\approx 1$.

\subsubsection{Comment on the acceptance}
\begin{figure}[!h]
    \centering
    \includegraphics[width=1\textwidth]{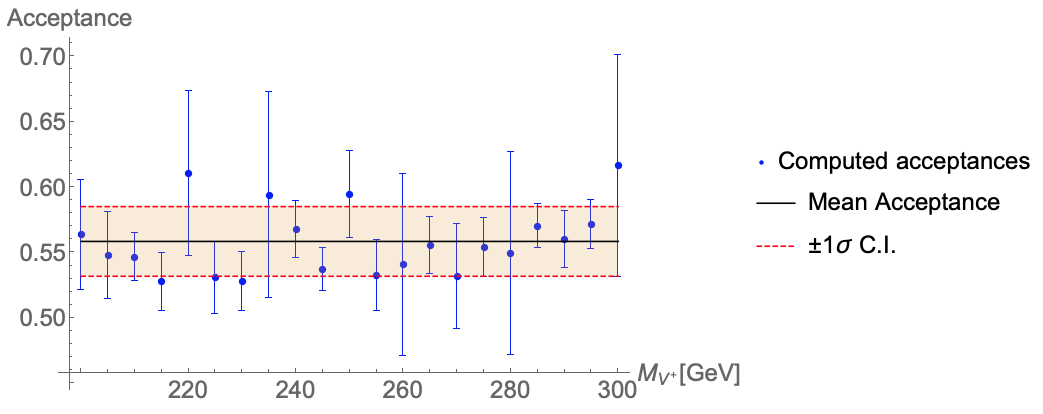}
    \caption{Acceptance for the simulated samples.}
    \label{accs}
\end{figure}

The calculation of the acceptance is straightforward. However, this quantity is sensitive to $N$, the total number of events in the sample. If we suppose that the number of events passing the cuts is a random variable obeying a Poisson distribution, the uncertainty of the acceptance can be computed as:
\begin{equation}
    \Delta_\mathcal{A}=\sqrt{\frac{\mathcal{A}}{N}}.
\end{equation}
Considering this, we can compute the acceptances for each sample, as can be seen in Figure \ref{accs}. Taking into account the error bars, the acceptance for the process tends to a constant value, which corresponds to the mean value of all the acceptances.

\subsubsection{Computation of the cross sections}
\begin{figure}[!h]
    \centering
    \includegraphics[scale=0.4]{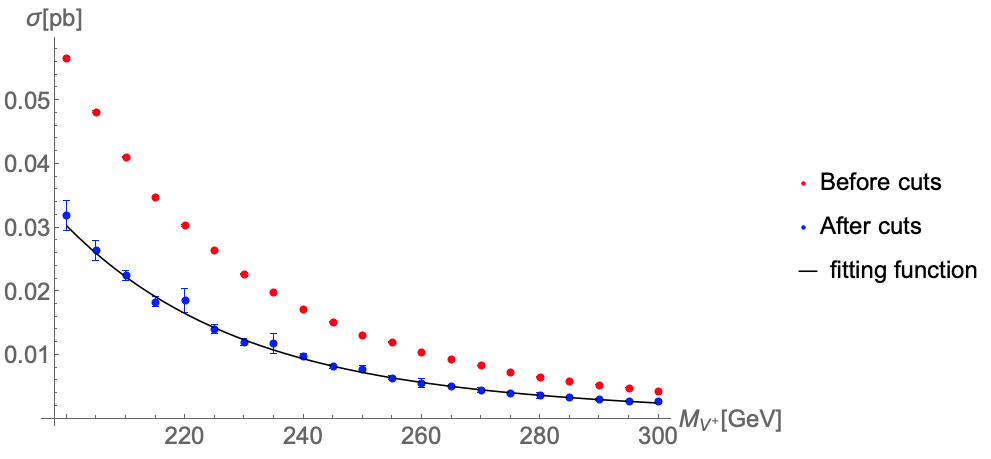}
    \caption{Results for the cross sections of the process before and after the cuts using $\beta_e=1$. The solid black line represents the fitting function over the scaled cross sections.}
    \label{deltam0sim}
\end{figure}

\begin{figure}[!h]
    \centering
    \includegraphics[width=0.5\textwidth]{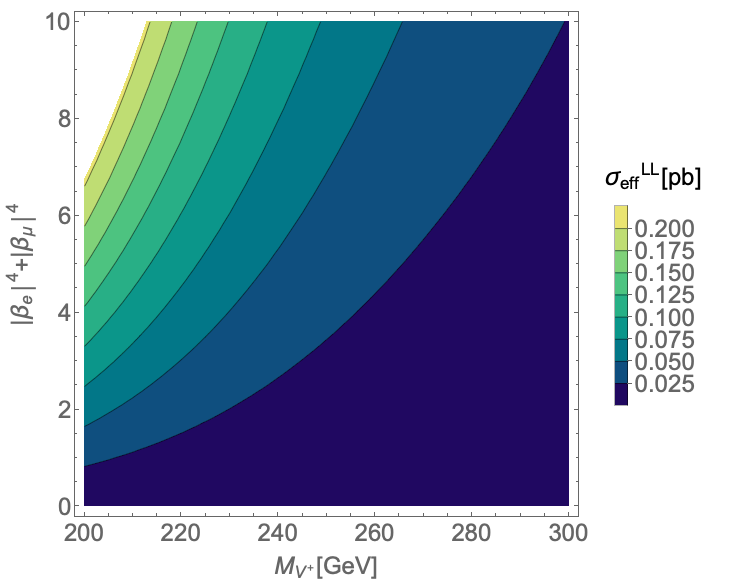}
    \caption{Parameter space and the corresponding values for the cross section of the process.}
    \label{scandeltam0}
\end{figure}

In order to absorb the acceptance fluctuations, we used the following fitting function to interpolate on values of the cross section as a function of $M_{V^+}$:
\begin{equation}
    f(x)=\sum_{n=0}^3 \frac{f_n}{x^n} .
\end{equation}
Where the fitting parameters can be found in Table \ref{paramsdeltam0}.
The fit considers the dependence on $M_{V^+}$ for $|\beta_e|^2=1$, as can be seen in Figure \ref{deltam0sim}. Taking this into account, the result can be extrapolated for different values of the coupling constant as:
\begin{equation}\label{stimcross}
    \sigma_{eff}^{ee}(|\beta_e|^2,M_{V^+})=|\beta_e|^4 f(M_{V^+}[\text{GeV}]).
\end{equation}
At this point it is relatively easy to include processes that include muons (instead of electrons) in the final step. First we observe that in our kinematical regime, the muon mass is negligible. Additionally, the topology of the processes involving muons is exactly the same to those involving electrons. Consequently, we can write a generalized version of eq. \eqref{stimcross}, and the total cross section for the process considering a muon or an electron lepton pair has the following form:
\begin{equation}\label{eqcross}
    \sigma_{eff}^{LL}=(|\beta_e|^4 +|\beta_\mu|^4) f(M_{V_{+}})
\end{equation}
In principle, a total cross section can also contain a $\tau$ lepton pair, but since these particles can decay hadronically it's reconstruction is more complicated, therefore we do not condider them. Besides that, the corresponding experimental reference considers only muons and electrons in the final state. A graphical representation of eq. \eqref{eqcross} can be seen in Figure \ref{scandeltam0}.
This expression allows to define bounds on the parameter space, considering the current searches on new physics. In order to define these bounds, we need to study the process kinematics.   
\begin{table}[!h]
    \centering
    \begin{tabular}{|c|c|c|}
    \hline
        fitting parameter & value & standard error  \\
        \hline
        $f_0$ & $-0.11$   &  $0.05$ \\
$f_1$& $97$& $36$\\
 $f_2$& $-3.0\times10^4$& $0.9\times10^4$\\
 $f_3$& $3.2\times10^6$& $0.8\times10^6$\\
 \hline
    \end{tabular}
    \caption{fitting parameters for $f(x)$}
    \label{paramsdeltam0}
\end{table}
 
\subsection{Kinematical Analysis without experimental constraints}\label{kinematical_analysis}
After completing the simulation, we studied the distribution of the relevant kinematical observables, which are defined and described in Table \ref{kinobjects}.

\begin{table}[!h]
    \centering
    \begin{tabular}{|c|c|}
    \hline
       object  & definition  \\
       \hline 
       $\cos\theta$  & angular distribution between the SFOS lepton pair\\
        $H_T$ & Scalar sum of the $p_T$ of all outgoing quarks.\\
       $M_{ee}$ &invariant mass of the SFOS lepton pair\\
       $p_T^{miss}$ & $p_T$ of the left handed HNL pair\\
       \hline
    \end{tabular}
    \caption{Relevant kinematical observables  for our study. We consider only an estimate of $H_T$ because we simulated the process at parton level.}
    \label{kinobjects}
\end{table}

For each simulated sample, we obtained the distribution of these variables. We also simulated the SM background, which is dominated by weak diboson production. We used Madgraph for the simulation of the following process: $p p \to j j e^+ e^- \nu \bar{\nu}$, where the neutrinos can be of any flavour. In order to gain an insight of the kinematical hints of the model, we studied the kinematics before and after applying the event selection criteria.
\subsubsection{Distributions without cuts}
\begin{figure}[!h]
\centering
    \includegraphics[width=0.8\textwidth]{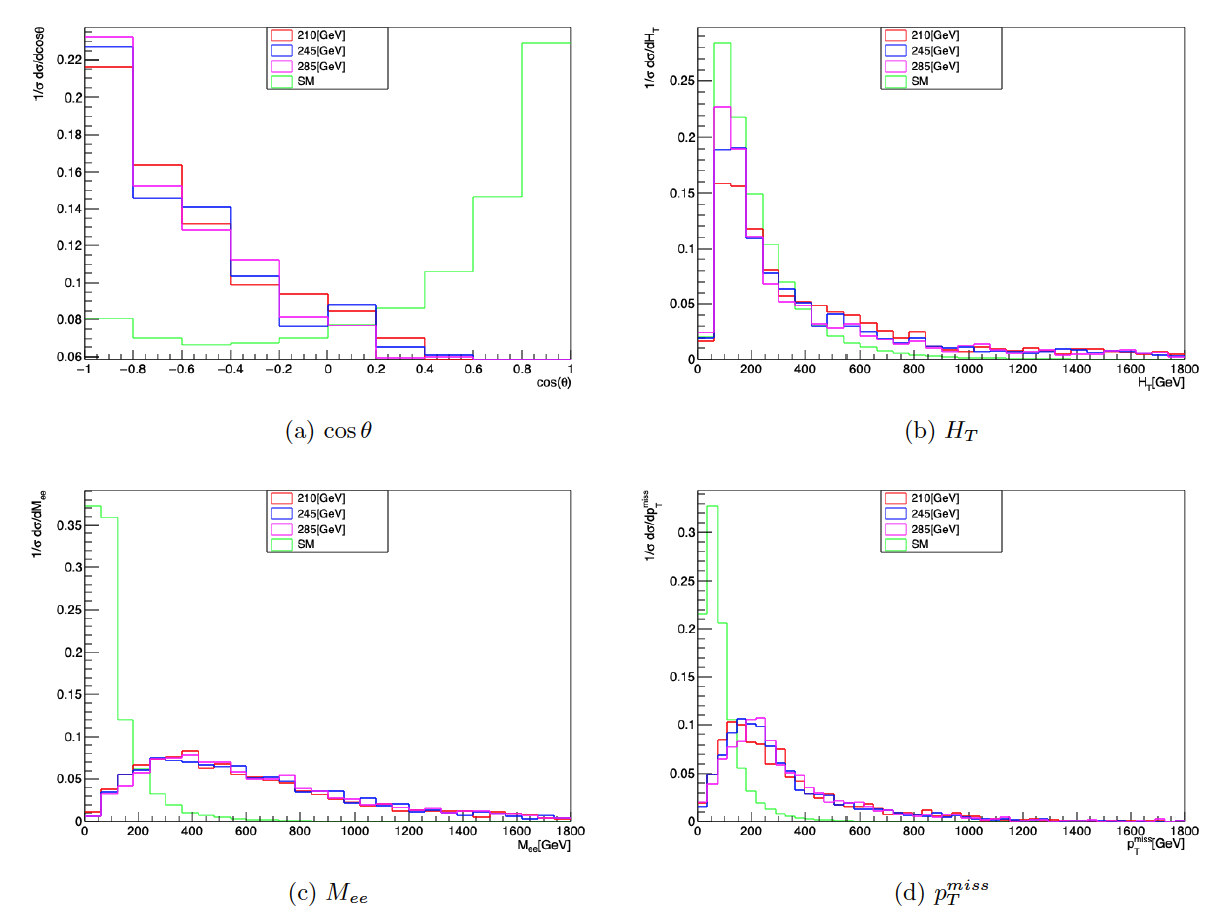}

\caption{Kinematical Distributions for some representative simulations without considering the event selection criteria}
\label{merged}
\end{figure}

A condensed histogram with some representative cases can be found in Figure \ref{merged}. 

All the signal samples have similar distributions, independent of the value of $M_{V^+}$. The strongest difference with the SM is the angular distribution of the lepton pair. On the other side, the $H_T$ distribution is practically the same for our signal and the background. Finally, $M_{ee}$ and $p_{T}^{miss}$ have some overlap between the signal and background distributions, however, in both cases the tail of the signal distributions is longer than the background distribution.  
\newline
With these considerations, we need to define a search strategy, which depends on the lifetime of the charged vectors.

\subsection{Lifetime of the Charged Vector}
\begin{figure}[!h]
    \centering
    \includegraphics[width=\textwidth]{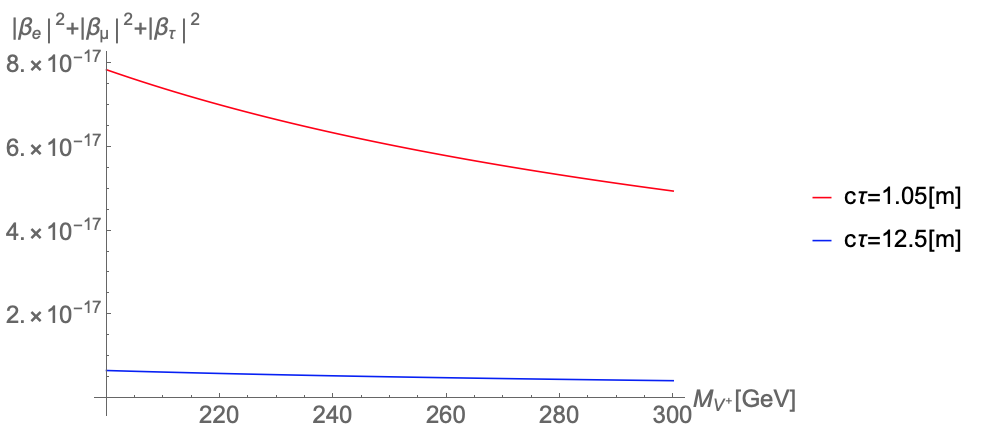}
    \caption{Parameter space with some characteristic decay widths.}
    \label{gammalims}
\end{figure}

The search strategy for the process depends on the Charged Vector lifetime. It's decay width was obtained with FeynRules, and, if we neglect the mass of the SM fermions, can be written as:
\begin{equation}
    \Gamma_{V^+}=(|\beta_e|^2 + |\beta_\mu|^2+|\beta_\tau|^2)\frac{ (M_{V^+}^2 - M_N^2)^2   (2 M_{V^+}^2 +M_N^2) }{(48 \pi M_{V^+}^5)}.
\end{equation}
The ATLAS dimensions are relevant quantities for particle reconstruction.  The Inner Detector of the ATLAS experiment has a diameter of 2.1[m], while the whole ATLAS detector has a diameter ATLAS detector has a diameter of 25[m] This information is publicly available in the ATLAS website (see Ref. \cite{atlas_disc}). This quantities are relevant because if the decay length of the charged Vector is in this range, the search strategy would be different, and we should look for Long Lived Particles (LLP). Nevertheless, the regions of the parameter space with these decay lengths have considerably small couplings (See Figure \ref{gammalims}). Since we are looking for upper bounds for higher values of the couplings, we can treat each outgoing particle as a prompt object. Therefore, the limits and methods applied in Ref. \cite{susylims} can be applied to our simulated samples.
\subsection{Experimental constraints from ATLAS searches}\label{colliderlims}
\begin{figure}[!h]
\centering
\includegraphics[width=\textwidth]{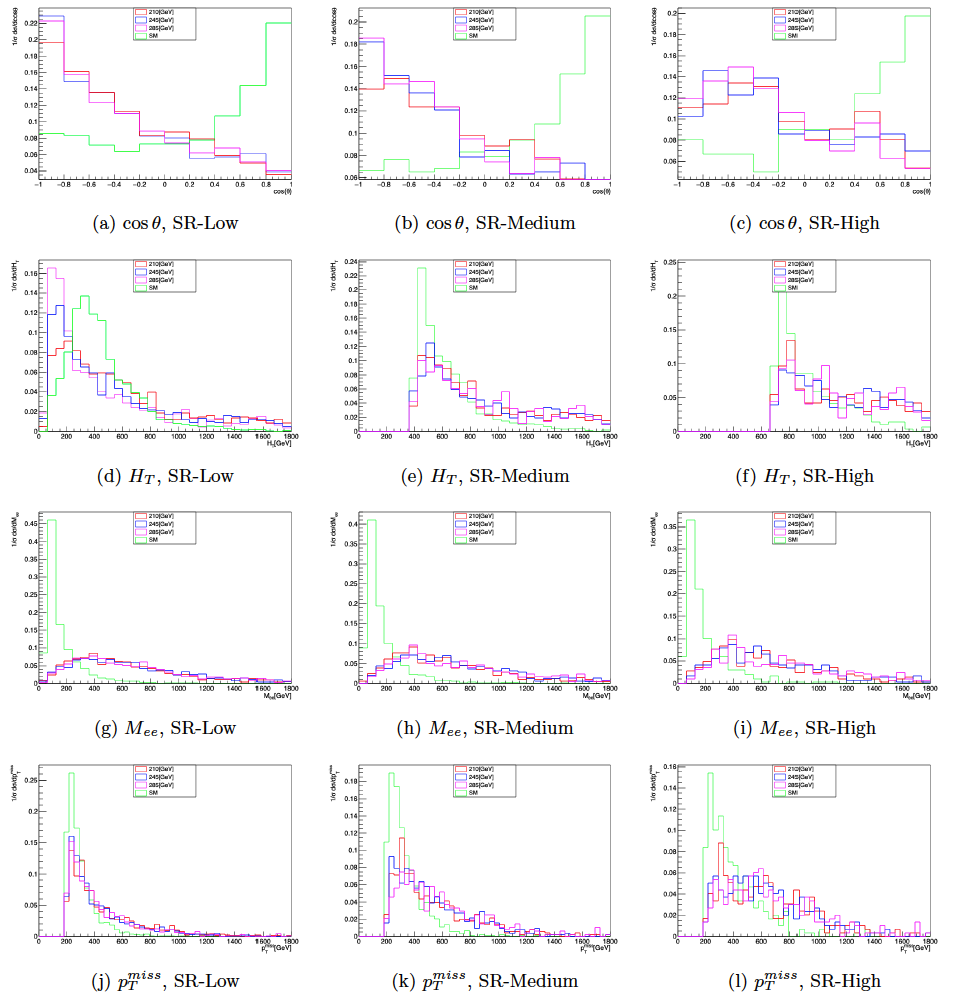}
\caption{Kinematical Distributions for some representative simulations after the application of the cuts.}
\label{cuts}
\end{figure}

 The resulting distributions after applying the ATLAS cuts can be found in Figure \ref{cuts}

As can be seen, the clear shape of the angular distribution is lost when the $H_T$ threshold increases. Since the angular distribution is the best observable to identify the signal, we must restrict our analysis to SR-Low. Additionally, the SM background can be severally removed by choosing a threshold on $M_{ee}$, but the choice of this threshold is non trivial due to the overlap between the distributions.
\subsubsection{Limit definition}
The background cross section reported in Ref. \cite{susylims}  has a value of $\sigma_{bkg}=102\pm 7$[fb] for a luminosity of $14.7$[fb $^{-1}$]. Since the reference didn't report a significant deviation from the Standard Model prediction, we can define a threshold using the significance of a measurement. To do so we used the standard definition of the significance $Z$, which has the following form:
\begin{equation}\label{sig}
    Z=\frac{s}{\sqrt{s+b}},
\end{equation}
where $s$ and $b$ are the number of signal and background events, respectively. Moreover, the number of events can be written in terms of the process cross section $\sigma$ and the collider luminosity $\mathcal{L}$: 
\begin{equation}\label{n=sigmaL}
    n=\sigma\mathcal{L}.
\end{equation}
We can rewrite eq. \eqref{sig} using eq. \eqref{n=sigmaL}, and use this result to constrain $\sigma_{eff}^{LL}$:
\begin{equation}
    \frac{\sigma_{eff}^{LL}\sqrt{\mathcal{L}}}{\sqrt{\sigma_{bkg}+\sigma_{eff}^{LL}}}\leq 1\implies \sigma_{eff}^{LL}\leq 2.67[\text{fb}].
\end{equation}

\begin{figure}[!h]

\centering
    \includegraphics[width=0.6\textwidth]{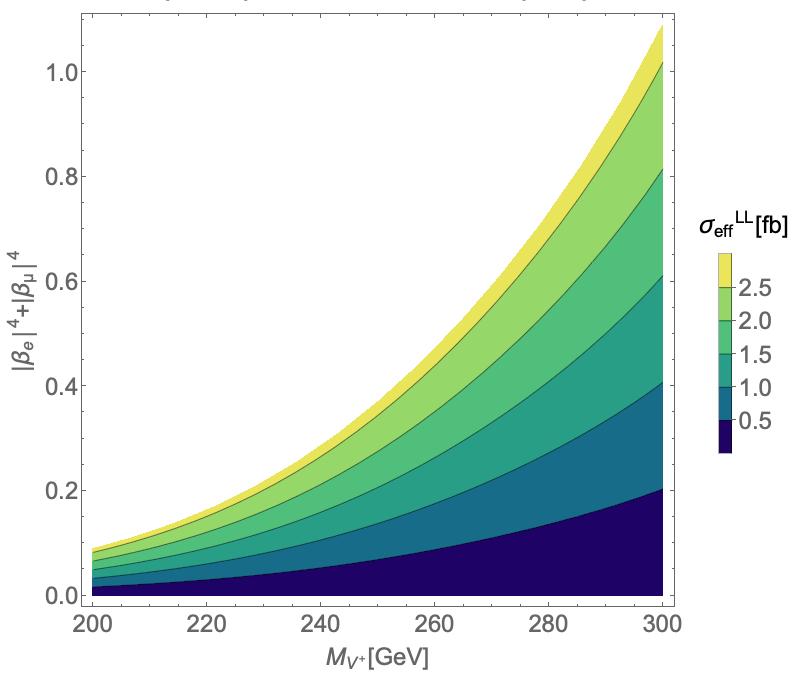}
    \caption{Available parameter space with the predicted cross sections. The white region is excluded.}
\label{paramlims}
\end{figure}

\begin{figure}[!h]
    \centering
    \includegraphics[width=0.5\textwidth]{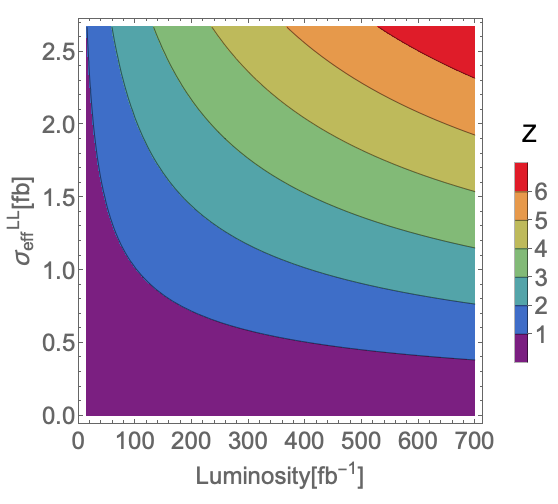}
    \caption{Significance for the measurement of the process as a function of the luminosity.}
    \label{lumilims}
\end{figure}

With this bound we can use \eqref{eqcross} to constrain the parameter space, as can be seen in Figure \ref{paramlims}. Also, it is possible to make predictions for the expected luminosity for a significant measurement of the process, as can be seen in Figure \ref{lumilims}. The predictions show that is not possible to surpass the discovery threshold within the expected luminosity of HL-LHC, but significances of $Z=3\sigma$ can be reached in the sensitivity range of current and future collider experiments. 

\section{Cut optimization}\label{cut_opt}
Up to now, we have shown that, under the current search strategies, the signal for our process is too weak to be detected in the upcoming years. In this section, we intend to find optimized cuts in order to increase the signal significance considering the short term projections for the luminosity. We applied two approaches, one based on a classical application of cuts and a second one based on machine learning techniques.
\subsection{Application of additional kinematical cuts}\label{traditional}
 In order to study the 
 effect of additional cuts on the signal significance,
 we considered the SM simulation (from now on, the background sample), which contains 46312 events. On the other side, we considered the simulated sample obtained for $M_{V^+}=245[$GeV$]$ (from now on, the signal sample), which contains 1946 events. The first attempt to separate the signal from the background is the application of cuts. We considered 4 cases, the first one with the cuts of the SR-low, which we will refer as the ATLAS cuts (AC for short), the other cases include additional cuts over $M_{ee}$. Since the samples have a different size, we define the $\alpha$ ratio as the number of events surviving the cuts over the number of total simulated events. Additionally, we compute the significance using equation \eqref{sig}, normalized by $Z_0=Z$(ATLAS cuts).
For each case we obtained the  Table \ref{samplecuts}

\begin{table}[!h]
    \centering
    \begin{tabular}{|c|c|c|c|c|}
    \hline
       Sample   & AC & AC+$M_{ee}>300$[GeV]& AC+$M_{ee}>400$[GeV]&AC+$M_{ee}>500$[GeV] \\
       \hline
        sig. & $1045$ & $847$ & $729$  &  $629$  \\
       
     bkg. & $2092$ & $269$ & $145$  &  $84$  \\
        $\alpha$(sig.) &$0.54$&$0.44$ &$0.38$ &$0.32$\\
        $\alpha$(bkg.) & $0.045$ & $0.0058$& $0.0031$  & $0.0018$\\
        $Z/Z_0$ & $1$ & $1.36$& $1.32$  & $1.26$\\
        \hline
    \end{tabular}
    \caption{Calculation of the $\alpha$ ratios for different cut settings.}
    \label{samplecuts}
\end{table}
We can see that the addition of the mass cuts improves the background rejection in one order of magnitude. Naturally, there is also a decrease in the signal events, due to the overlap between the distributions. The signal rejection produces a decrease in the relative significance for bigger $M_{ee}$ thresholds. 
Since the optimal threshold choice is not trivial, we applied an iterative process to find the optimal threshold.

\subsubsection{The iterative process}
\begin{figure}[!h]
    \centering
    \includegraphics[width=0.5\textwidth]{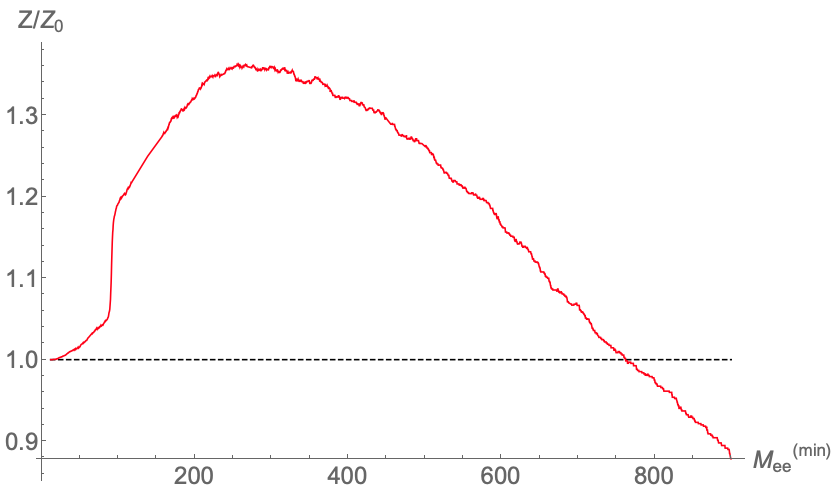}
    \caption{Relative significance as a function of t $M_{ee}^{(min)}$.}
    \label{mthre}
\end{figure}

The algorithm used was based on the iterative application of the event selection cutflow adding an additional cut of the form: $M_{ee}>M_{ee}^{min}$, where $M_{ee}^{min}$ is a dynamical variable that takes every value in the range of $M_{ee}$. For each iteration, we computed the relative significance.
From now on, we will refer to this process as the application of dynamical thresholds. 
\newline
 As can be seen in Figure \ref{mthre}, The relative significance has a peak for a threshold of $M_{ee}^{(min)}\approx 260$[GeV] with a value of $Z/Z_0\approx 1.36$. 

We performed the same analysis for the missing energy and the dilepton angle, defining the cuts as follows:
\begin{itemize}
    \item $p_{T}^{miss}>p_{T}^{miss(min)}$
    \item $\cos(\theta)< \cos\theta^{(max)}$
\end{itemize}

\begin{figure}[!h]
    \centering
    \includegraphics[width=0.5\textwidth]{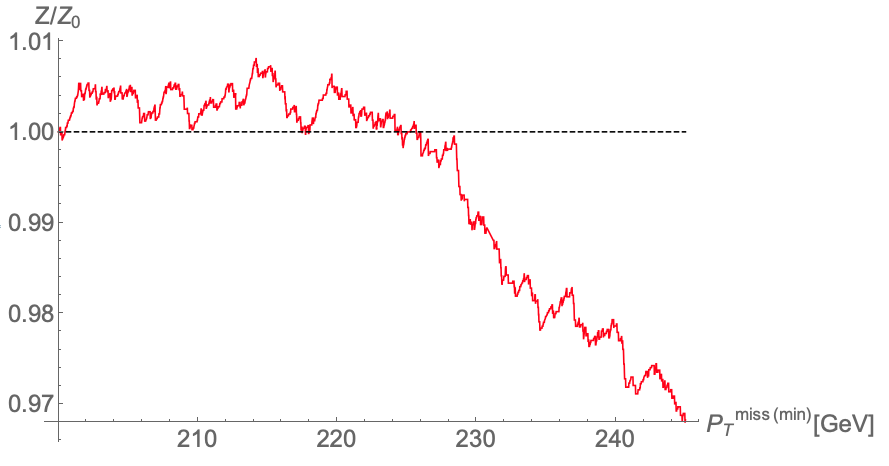}
    \caption{Relative significance as a function of  $p_{T}^{miss(min)}$.}
    \label{ptmissthre}
\end{figure}

\begin{figure}[!h]
    \centering
    \includegraphics[width=0.5\textwidth]{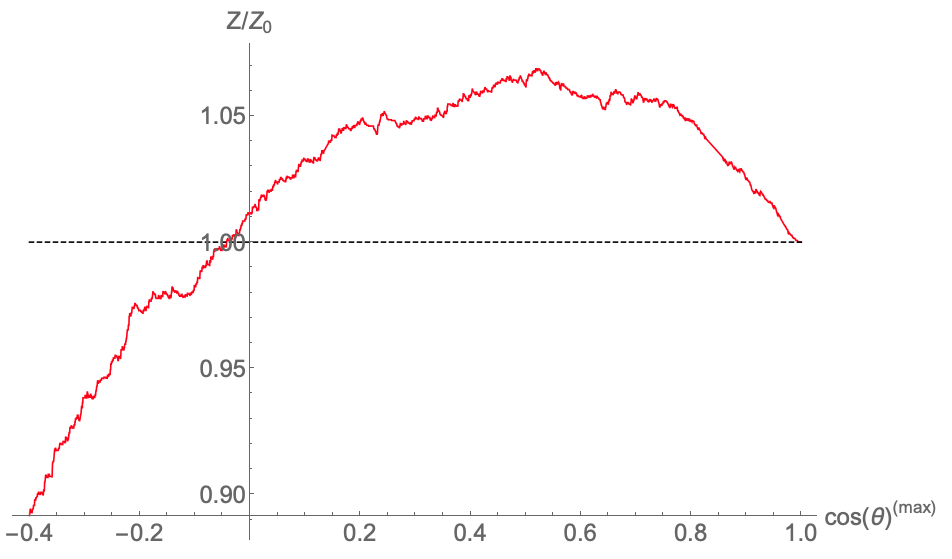}
    \caption{Relative significance as a function of  $\cos\theta^{(max)}$. }
    \label{cthetathre}
\end{figure}

Note that the cut on the angular distribution is an upper bound, instead of the lower bounds used in $M_{ee}$ and $p_{T}^{miss}$.
The relative significance as a function of the new thresholds
can be seen in Figures \ref{ptmissthre} and \ref{cthetathre}. In these cases, the significance enhancement is lower.


\subsubsection{Application of two cuts at the same time}
\begin{figure}[!h]
    \centering
    \includegraphics[width=\textwidth]{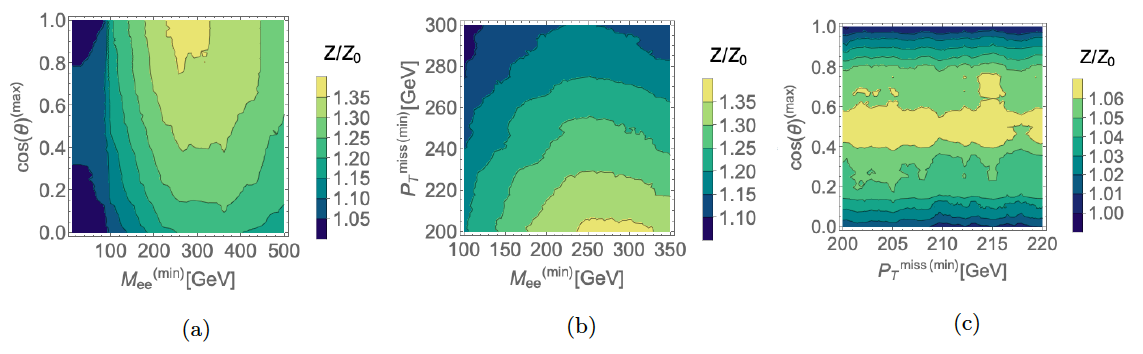}
    \caption{Relative significance for applying two cuts at the same time.}
    \label{twocutsmerg}
\end{figure}

For the sake of completeness, the application of two dynamical threhsolds at the same time can be seen in Figure \ref{twocutsmerg}. There is no significant gain in the significance enhancement, and the mass cut by itself gives the best result.

\begin{figure}[!h]
    \centering
    \includegraphics[width=\textwidth]{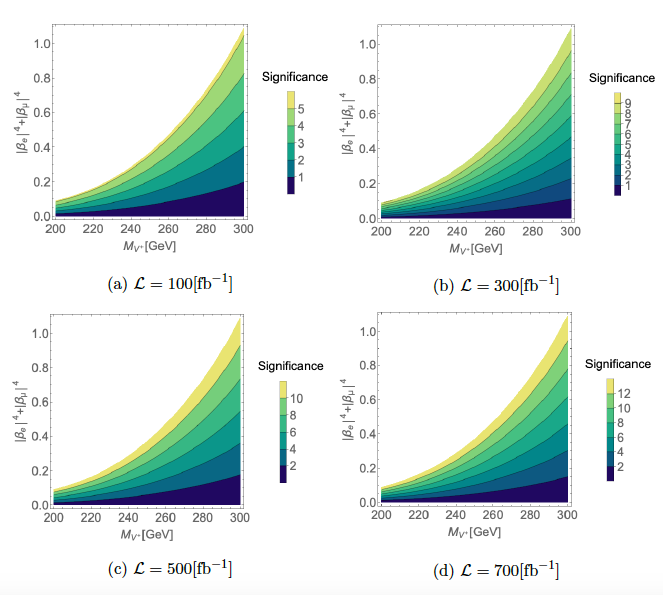}
    \caption{Significance projections for different values of the luminosity, considering the mass cut.}
    \label{atlas_vs_opt}
\end{figure}

Therefore, the only additional cut we impose is $M_{ee}>260[$GeV$]$. With this we can compute the signficance enhancement and obtain the signal projections for different values of the luminosity, as can be seen in Figure \ref{atlas_vs_opt}.

Despite the successful results obtained for the significance enhancement, the dynamical threshold algorithm can be slow and computationally demanding for events with a larger number of relevant kinematic objects. Therefore, in addition to the result obtained in this section, we tried a different approach based on novel techniques used in the area of data analysis, based on machine learning.

\subsection{Significance optimization via Machine Learning Techniques}
In the previous section, we applied additional cuts on the kinematical variables, which can be understood as a decision boundary in the space of kinematical variables (from now on, the feature space). The decision boundary obtained by cuts corresponds to straight lines in the feature space. This approach is not optimal when there are correlations between the different kinematical variables. Nevertheless, these correlations are not trivial and difficult to identify. Therefore, to find a flexible decision boundary it's necessary to apply machine learning (ML) techniques for feature space studies. 
\subsubsection{Feature space definition}
\begin{figure}[!h]
    \centering
    \includegraphics[width=0.55\textwidth]{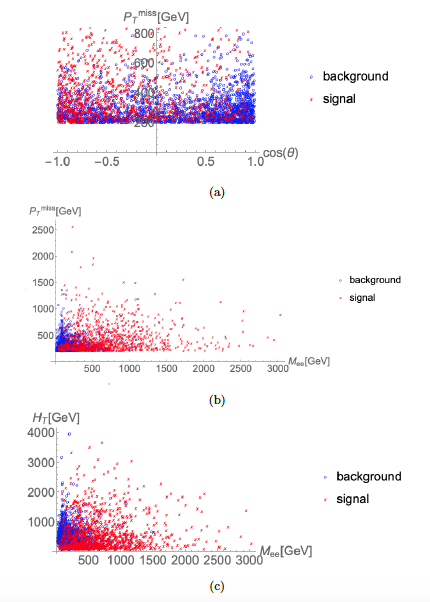}
    \caption{Some projections of the feature space}
    \label{feat_space}
\end{figure}

We consider the signal and background samples defined in Section \ref{traditional}, after applying the ATLAS cuts. A point in the feature space is defined as follows:

\begin{equation}
    \overrightarrow{\mathcal{F}}= \{M_{ee},p_T^{miss},\cos\theta,H_T\}
\end{equation}
Some projections of the feature space can be see in Figure \ref{feat_space}

\subsubsection{ML implementation}
\begin{figure}[!h]
    \centering
    \includegraphics[width=0.5\textwidth]{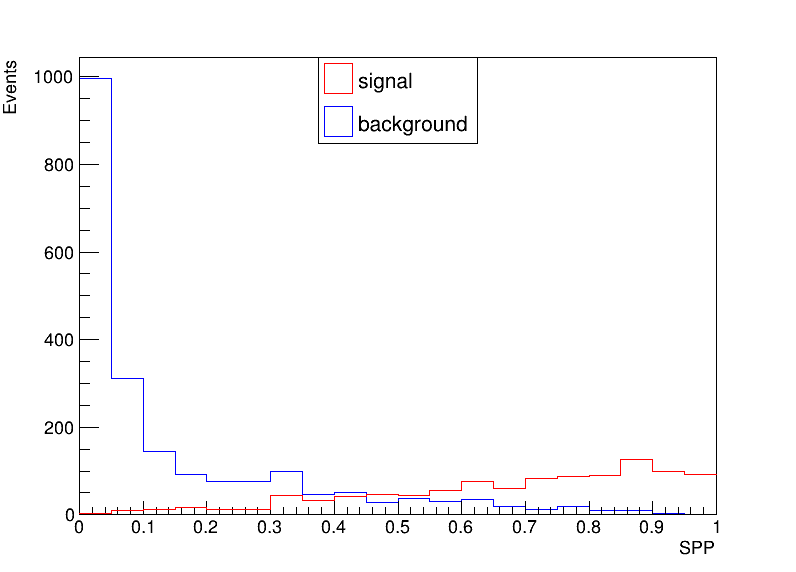}
    \caption{SPP for simulation samples, considering both training and testing datasets.}
    \label{spp}
\end{figure}

\begin{figure}[!h]
    \centering
    \includegraphics[width=0.7\textwidth]{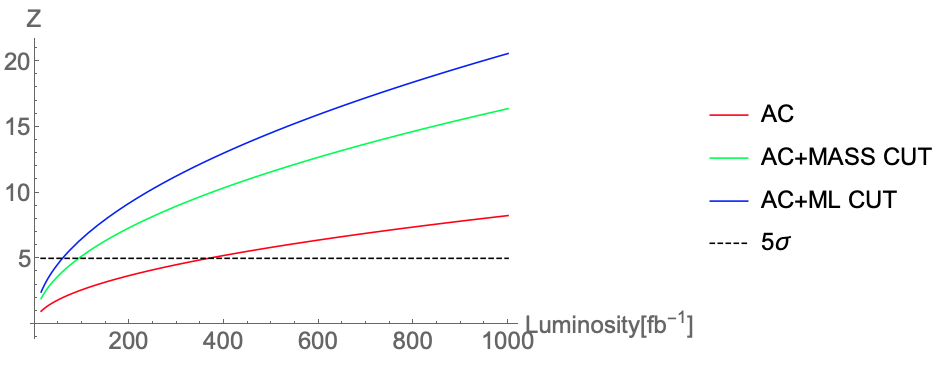}
    \caption{Significance benchmark for the upper bound on the parameter space.}
    \label{sig_bench}
\end{figure}

\begin{figure}[!h]
    \centering
    \includegraphics[width=0.6\textwidth]{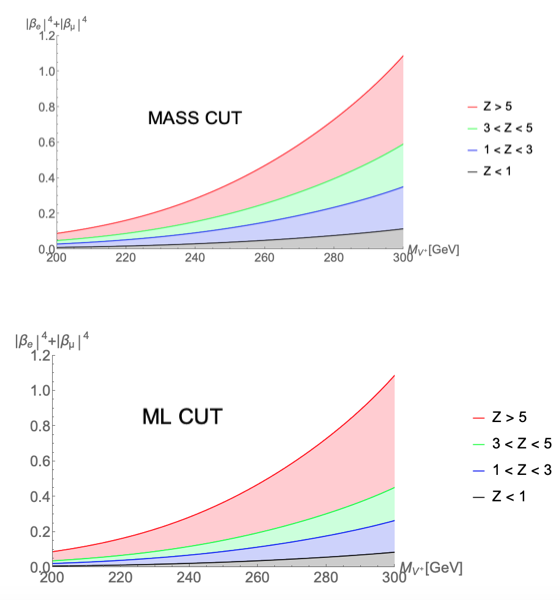}
    \caption{Parameter space mapping considering the predicted significance for a luminosity of $300$[fb$^{-1}$]}
    \label{proj_params}
\end{figure}

We used Mathematica to train a classifier that can identify a point in the feature space as signal or background, giving as an output the probability of a point to be labelled as signal. We considered 800 events from the signal sample and the first 1600 events from the background sample to train the classifier. We used the \emph{Classify} Mathematica function, which computes several classification algorithms and chooses the optimal for the task. The best result was obtained with an algorithm based on nearest neighbors, which uses the information of the closest points in the feature space to make the prediction. We used the remaining events to test the behaviour of the classifier. The confusion matrix associated to the test sample can be seen in Table \ref{confussion}. Additionally, we calculated the probability of an event for being classified as signal (SPP for short) for the complete sample, as can be seen in Figure \ref{spp}. A large part of the background can be removed by the classifier, but there is still a relevant overlap that must be taken into account. To perform a benchmark with the standard cut flow of Section \ref{cut_opt}, we computed the relative significance considering a cut of the form $SPP\geq 0.5$, obtaining:

\begin{equation}
    \frac{Z^{ML}}{Z_0}=1.39
\end{equation}
This value is slightly bigger than the significance enhancement obtained in the previous Section, but has the problem of being model dependent. An additional mass cut can be suited into different searches for New Physics related with heavy mediators, but the classifier is made specifically for this model. Due to the data-driven logic of the classifier, this cut lacks of physical meaning. Therefore, it's application on other models could be not accurate.  For the sake of completeness, we computed a benchmark for the significance projections considering only AC, AC with the mass cut and AC with the ML cut, as can be seen in Figure \ref{sig_bench}. Additionally, computed the scale factors that correct the cross section after the application of the new cuts, as can be seen in Table \ref{SFs}. 
Finally, the predicted significance for a luminosity of $300[\text{fb}^{-1}]$ can be written in terms of the parameter space, in order to facilitate the parameter exclusion when the measurements  are performed (See Figure \ref{proj_params}).

\begin{table}[!h]
    \centering
    \begin{tabular}{|c|c|c|c|}
    \hline
         & Predicted Signal & Predicted background &  \\
         \hline
     Real Signal    & 454 & 38 & 492 \\
     Real Background    &54 &191 &245  \\
     \hline
        &508 &229 &  \\
        \hline
    \end{tabular}
    
    \caption{confusion matrix for the test sample}
    \label{confussion}
\end{table}
\begin{table}[!h]
    \centering
    \begin{tabular}{|c|c|c|c|}
       \hline
      cut   & signal SF & background SF &Maximum number of events ($\mathcal{L}=300[\text{fb}^{-1}]$) \\
      \hline
    Mass cut     & $0.86$ &$0.17$&689\\
    ML cut & $0.78 $ & $0.08$&625\\
    \hline
    \end{tabular}
    \caption{Benchmark of the different approaches for cut optimization.}
    \label{SFs}
\end{table}

\subsubsection{Decision boundary comparison}
\begin{figure}[!h]
    \centering    \includegraphics[width=\textwidth]{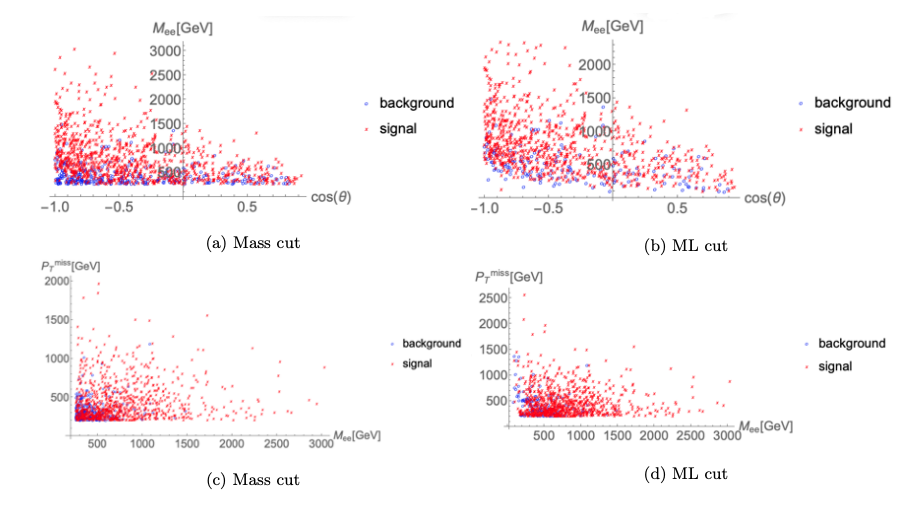}
    \caption{Projections on the feature space after applying the two different cuts.}
    \label{decisions}
\end{figure}

After applying the different proposed cuts, the feature space changes. In Figure \ref{decisions} we can see the effect of the mass cut as a straight line in the feature space, while the ML cut generates smoother decision boundary.  

\section{Further calculations}

\subsection{Effect of gauge-like coupling in the production cross section}
In this section, we study the effect of the gauge-like coupling proposed by the authors of Ref. \cite{vietnam} in our result. To do so, we simulated the process u u $\to $ u u $N_L$ $N_L$ $e^+$ $e^-$ in madgraph, for different values of $1/\xi$, as can be seen in Table \ref{benchmark}. For these simulations, we fixed the vector mass to $M_{V^+}=250$[GeV].
\begin{table}[!h]
    \centering
    \begin{tabular}{|c|c|}
    \hline
      $1/\xi$   &  cross section[pb] \\
      \hline
    0    &$ 0.001379 \pm 4.8\times10^{-6}$\\
     0.001    &$ 0.001373 \pm 7.6\times10^{-6}
$\\
     0.01    &$ 	
0.001791 \pm 7.4\times10^{-6}
$\\
     0.1    &$1.098 \pm 0.0027 $\\
     1    &$1.004\times10^{4} \pm 30 $\\
     \hline
    \end{tabular}
    \caption{Benchmark for the production cross section for different values of $1/\xi$}
    \label{benchmark}
\end{table}
We can see that cross section increases with the value of $1/\xi$, with a strong jump for $1/\xi \geq 0.1$, obtaining  an abnormally big cross section for $1/\xi=1$. The  observed behaviour cannot be explained by means of the numerical simulations, therefore the effect of non minimal gauge interactions must be studied in more depth, but these studies are beyond the scope of this work. 
\subsubsection{Kinematical benchmark}
\begin{figure}[!h]
    \centering
    \includegraphics[width=\textwidth]{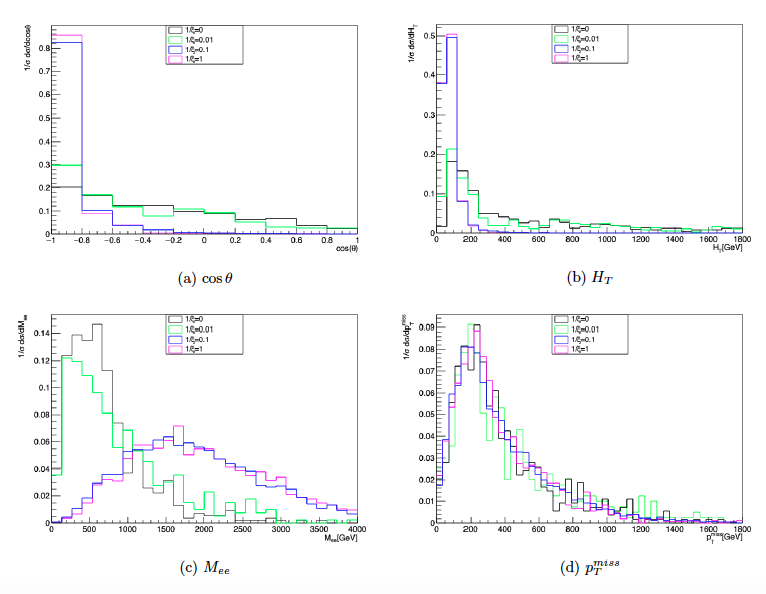}
    \caption{Knematical distributions for different values of $1/\xi$.}
    \label{bench}
\end{figure}

We obtained the relevant observables for the generated samples without any selection criteria, as can be seen in Figure \ref{bench}. We can see that, at the qualitative level, the distributions are similar, except for the dilepton invariant mass, where the peak is displaced for the highest values of $1/\xi$. Therefore, the search strategy and the cuts can be applied, but the limits on the parameter space will depend on $1/\xi$ and could change. 

\subsection{Considering higher HNL masses}
\begin{figure}[!h]
    \centering
    \includegraphics[width=\textwidth]{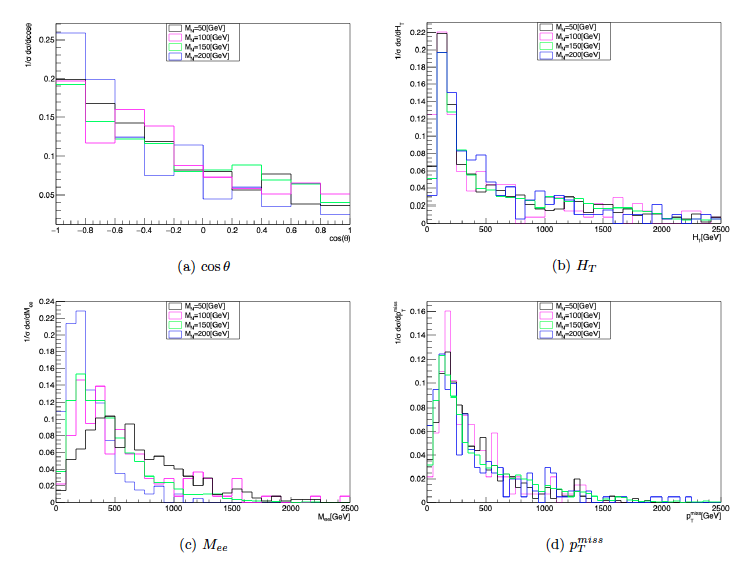}
    \caption{Kinematical distributions for different values of $M_N$.}
    \label{massbenchfig}
\end{figure}

The analysis performed in this work considered a fixed HNL mass of $50$[GeV]. In this section, we simulated events for the process $uu \to u u N_L N_L e^+ e^-$ considering different HNL masses and a fixed mass for the charged vector of $M_{V^+}=250$[GeV]. As can be seen in Table \ref{massbenchtab}, the cross section is not affected. This behaviour is explained because the cross section is determined by the production of the vector states, while the HNLs are decay products of these vectors. 
\newline 
In addition to the computation of the cross sections, we studied the kinematics of the process without any selection criteria. As can be seen in Figure \ref{massbenchfig}, the $M_{ee}$ distribution presents a shorter tail for higher HNL masses. However, these tails are larger than the SM background. For the rest of the variables, the kinematical distributions don't present significant differences for the different $M_N$ values.

\begin{table}[!h]
    \centering
    \begin{tabular}{|c|c|}
    \hline
      $M_{N}$[GeV]   & cross section [pb] \\
      \hline
       $50$  &    $0.001379 \pm 4.8\times 10^{-6}$    \\
       $100$  &   	
$0.001392 \pm 5.5\times 10^{-6} $    \\
       $150$ &   	
$0.001408 \pm 4.9\times 10^{-6}$
      \\
       $200$ &  	
$0.00136 \pm 1.4\times10^{-5}$       \\
\hline
    \end{tabular}
    \caption{Cross section for the process $u u\to u u N_L N_L e^+ e^-$ with $M_{V^+}=250\text{[GeV]}$}
    \label{massbenchtab}
\end{table}

\subsection{Annihilation cross section}\label{sec:lowerlims}
\begin{figure}[!h]
    \centering
    \includegraphics[width=0.6\textwidth]{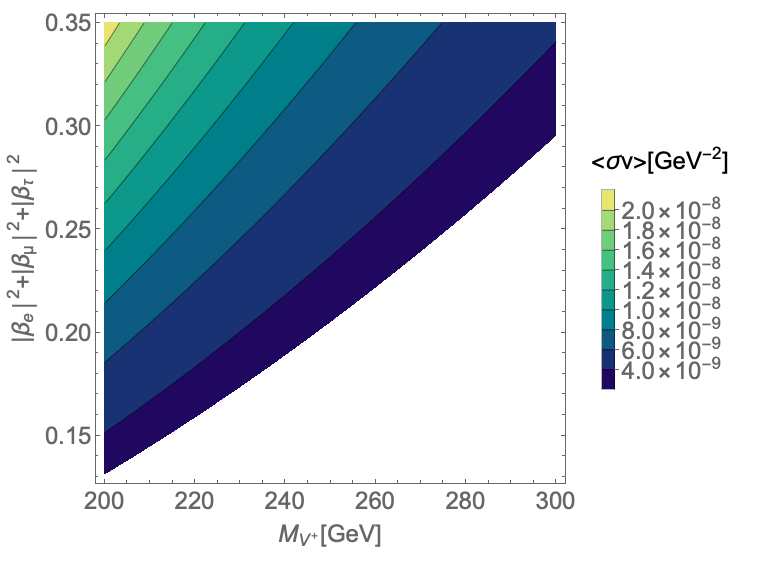}
    \caption{Annihilation cross section for the HNL in the region of interest. The white region is excluded by overabundance.}
    \label{dmlims}
\end{figure}

\begin{figure}[!h]
    \centering
    \includegraphics[width=0.7\textwidth]{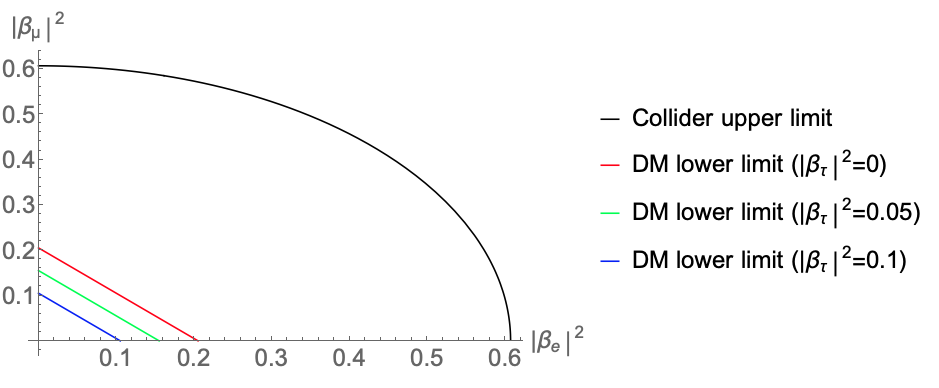}
    \caption{Limits for the ($|\beta_e|^2$,$|\beta_\mu|^2$) plane, for $M_{V^+}=250[\text{GeV}]$}
    \label{comblims}
\end{figure}

We can use eq. \eqref{annicross} to calculate the annihilation cross section for the region of interest. In order to avoid overabudance, this quantity must satisfy the following lower bound  (see Refs. \cite{langacker,pdg2022,Profumo}):
\begin{equation}
    \langle \sigma v\rangle\geq 3\times10^{-9}[\text{GeV}^{-2}].
\end{equation}
The application of this lower bound in the region of interest can be seen in Figure \ref{dmlims}.
The collider limits presented in Section \ref{colliderlims}  can be combined with the DM limits, as can be seen in Figure \ref{comblims}. 
The lower limit in the coupling space depends on the value of $|\beta_\tau|^2$, which cannot be constrained with the collider constraints presented in this work. However, there are several ways to constrain this parameter. An interesting possibility could be to consider radiative processes involving Lepton Flavor Violation (LFV), but these calculations are subtle due to the non minimal gauge structure of the model.   
\subsection{Revisiting the vector's lifetime}
\begin{figure}[!h]
    \centering
    \includegraphics[width=0.6\textwidth]{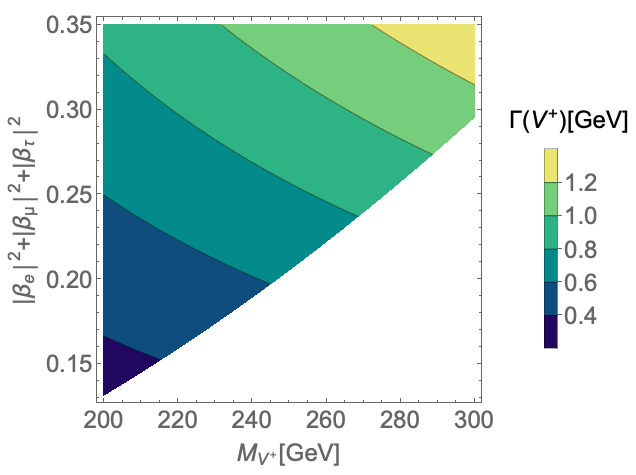}
    \caption{Allowed values for $\Gamma(V^+)$ considering the DM constraint}
    \label{bounded_gamma}
\end{figure}

The dark matter constraint can be used to find the minimum value of the charged vector's decay width, as can be seen in Figure \ref{bounded_gamma}. The minimum value for the coupling sum is achieved when $V^+=200$[GeV], with a value of $(|\beta_e|^2+|\beta_\mu|^2+|\beta_\tau|^2)_{min}=0.13$.  These restrictions can be used to estimate the maximum value for the decay length:
\begin{equation}
    c\tau^{max}\approx6.2\times 10^{-16}[m].
\end{equation}
The upper bound on the vector decay length ensures that the charged vector cannot behave as an LLP in the parameter space region considered in this work.
\section{Parameter space exploration for DM studies}
As we stated before, the model has two DM candidates. We want to focus on the scenario where the HNL is the DM candidate. In this section, we present some guidelines for DM phenomenology in the context of our model.

\subsection{Relic density}
\begin{figure}[!h]
    \centering
    \includegraphics[width=0.8\textwidth]{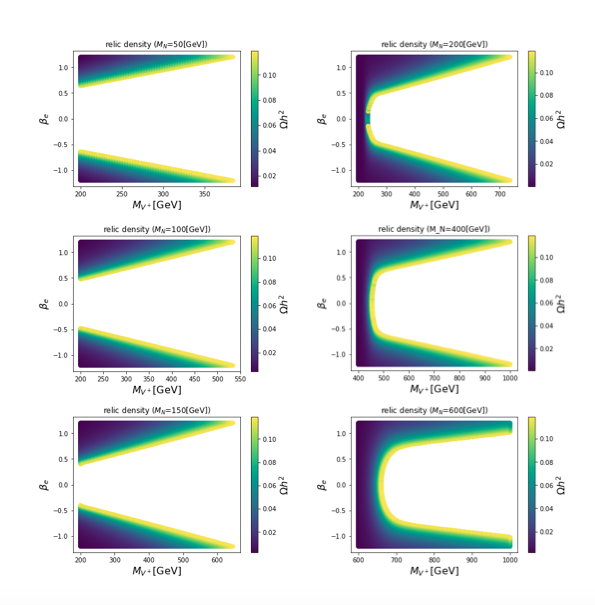}
    \caption{Relic density map in the parameter space for some characteristic values for $M_N$.}
    \label{relics}
\end{figure}

\begin{figure}[!h]
    \centering
    \includegraphics[width=0.6\textwidth]{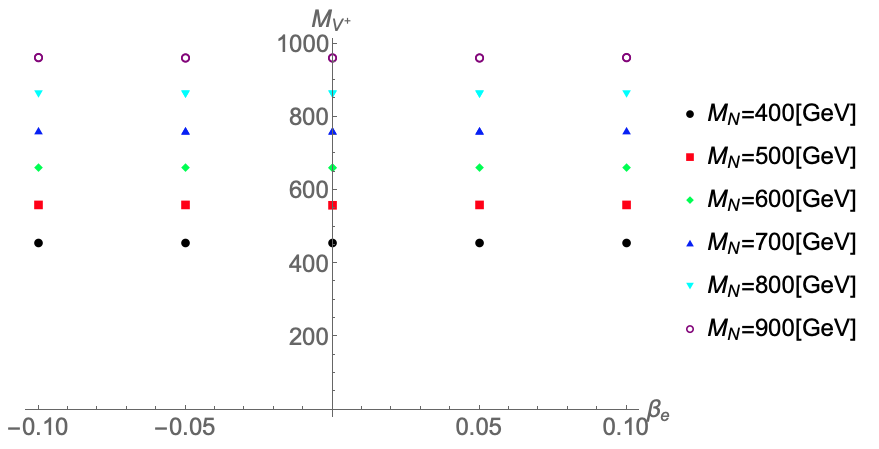}
    \caption{Plateau strcuture for different values of $M_N$.}
    \label{plateaus}
\end{figure}

\begin{figure}[!h]
    \centering
    \includegraphics[width=0.8\textwidth]{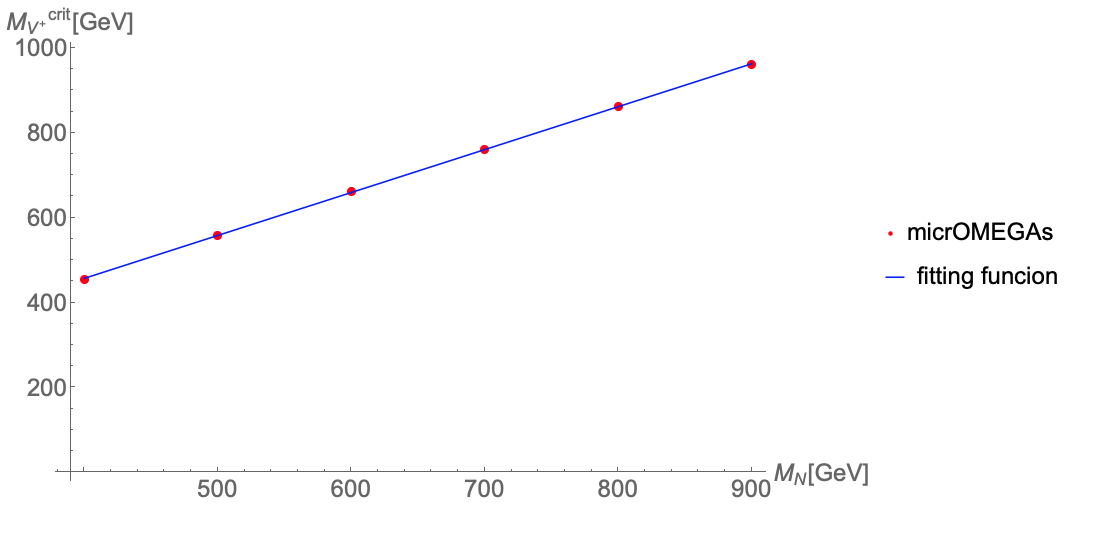}
    \caption{$M_{V^+}^{crit}$ dependence on $M_N$}
    \label{mvcrit}
\end{figure}

As was shown in Section \ref{sec:lowerlims}, there is a lower bound coming from dark matter constraints. However, the current lower limit is an approximation based on the annihilation cross section. A more rigorous limit must come from solving the Boltzmann equation. To do so, we restricted to the kinematical region where $\text{max}(M_N,200[\text{GeV}])\leq M_{V^+}\leq 1000$[GeV]. The lower limit has two conditions, the first condition comes from assuring that the HNL is the DM candidate, while the region $M_{V^+}<200$[GeV] is excluded by the results presented in Ref \cite{vector_dm}.
Moreover, we considered a special case where $\beta_\mu=\beta_\tau=0$, in order to avoid lepton flavor violating DM annihilation. With these considerations, we used micrOMEGAs \cite{micromegas1,micromegas11,micromegas2,micromegas3} to obtain the HNL relic density for different regions of the parameter space, as can be seen in Figure \ref{relics}.  
In this Figure, we excluded the points with $\Omega h^2\geq 0.12$, observing the expected lower bound for the coupling. It's worth to mention the relaxation on this lower bound for higher $M_N$ values. Moreover, there is a tendency for a plateau in the relic density as a function of $\beta_e$ when $M_N$ and $M_{V^+}$ have comparable values. In this region, the vector doublet decay is suppressed and these particles are stable enough to contribute to the thermally averaged cross section.
In order to study the plateau formation, we used micrOMEGAs for making a parameter scan considering $|\beta_e|\leq0.1$ and $M_N<M_{V^+}\leq1000$[GeV] for different values of $M_N$. After completing the scan, we made a point selection based on the condition $0.11<\Omega h^2 <0.12$. The scan result can be seen in Figure \ref{plateaus}. In this region, the relic density doesn't depend on $\beta_e$, and there is a critical mass $M_{V^+}^{crit}$ that saturates the relic density for each value of $M_N$. The relationship between $M_N$ and $M_{V^+}^{crit}$ is linear and described by the following function:
\begin{equation}\label{eq:mvcrit}
    M_{V^+}^{crit}(M_N)=a M_N+ b,
\end{equation}
where the fitting parameters have the following values: $a=1.011\pm0.004$ and $b=53\pm 3$. This result can be seen in Figure \ref{mvcrit}. It's worth to mention that the plateau structure depends on the gauge structure of the model, and eq. \eqref{mvcrit} might change for a different parameter choice. Naturally, this dependence is less relevant when the leptonic interaction dominates the annihilation cross section.

\begin{figure}[!h]
    \centering
    \includegraphics[width=0.8\textwidth]{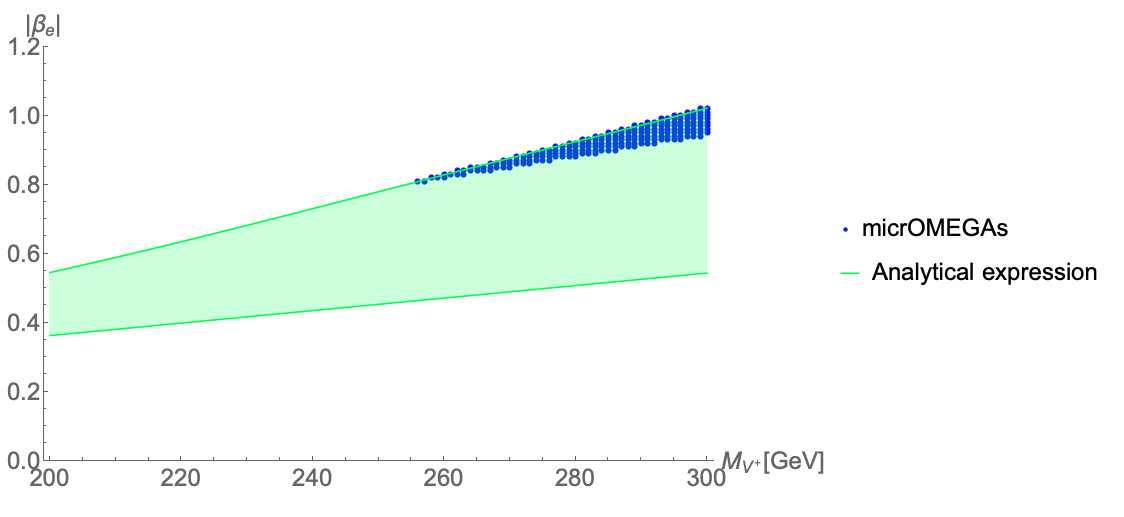}
    \caption{Available parameter space for $M_N=50\text{[GeV]}$.}
    \label{strict}
\end{figure}

For the sake of completeness, we combined the upper limits for the region of interest for collider searches for $M_N=50$[GeV]. The relic constraint is severally more strict when the HNL couples only to electrons, as can be seen in Figure \ref{strict}. These limits are very severe due to the $\beta_\mu=\beta_\tau=0$ assumption. As we stated in Section \ref{sec:lowerlims}, these limits are senstive to the couplings between the dark sector and the SM fermions, and  should be confronted with LFV calculations.

\subsection{Direct detection and electronic recoil}
\begin{figure}[!h]
    \centering
    \includegraphics[width=0.4\textwidth]{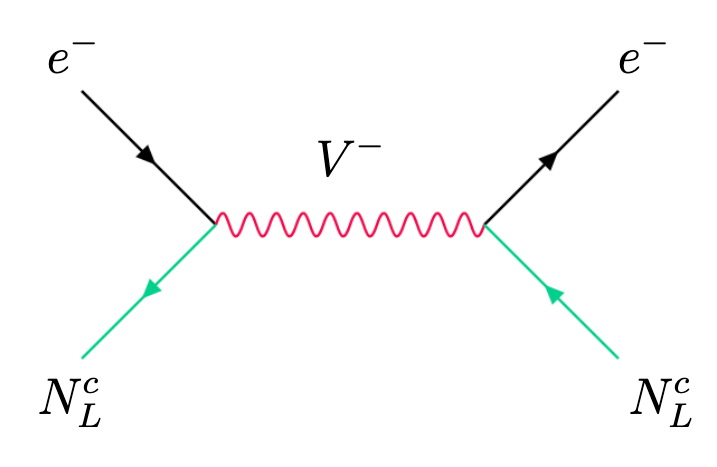}
    \caption{Elastic scattering between DM and electrons in the context of our model.}
    \label{elastic_scat}
\end{figure}

Direct detection experiments have shown an excess of electronic recoils that could be explained by interactions between dark matter and electrons (Ref. \cite{directdetection}). Interactions of this kind naturally appear in our model. The elastic scattering $N_L e^- \to N_L e^-$ has a contribution from the diagram depicted in Figure \ref{elastic_scat}. The corresponding squared amplitude for this process has the following form:
\begin{equation}
\begin{split}
    \bar{|\mathcal{M}|}^2&=\frac{|\beta_e|^4}{4 M_{V^+}^4 ((M_{V^+}^2-s)^2+M_{V^+}^2 \Gamma_{M_{V^+}}^2)}\big[
    m_e^2 (-2 M_N^2 (4 M_{V^+}^4+s^2)+M_N^4 (2 M_{V^+}^2-s)+\\
    &s (s-2 M_{V^+}^2)^2+8 M_{V^+}^4 t)+m_e^4 (2 M_{V^+}^2-s) (M_N^2-2 M_{V^+}^2+2 s)+m_e^6 (s-2 M_{V^+}^2)+\\&M_N^2 (4 M_{V^+}^4 (s+2 t)+s^3)+M_N^6 (s-2 M_{V^+}^2)+2 M_N^4 M_{V^+}^2 (s-2 M_{V^+}^2)\\&-4 M_{V^+}^4 t (s+t)
    \big].
    \end{split}
\end{equation}
However, the elastic scattering between DM particles and electrons is more complex, due to the fact that electrons are bounded by the atom nucleus. The calculation of this scattering needs the consideration of form factors that can account for the nucleus effects. Besides, the excess in electronic recoil is relevant for low DM mass, which is not the case considered in this work. However, the low mass region could be studied in future works and this expression may be useful for those calculations.

\chapter{Conclusions and projections}

In this work, we studied an extension of the standard model based on a massive $SU(2)_L$ vector doublet and a left-handed heavy neutral lepton. After showing the theoretical motivations supporting this model, we studied the model phenomenology in the context of the large hadron collider. We showed how the model can produce a signal characterized by an SFOS lepton pair, jets and missing energy. In this final state, the model predicts a large number of back to back leptons and a large value for the lepton pair invariant mass. Moreover, we defined an upper bound on the parameter space
based on current searches for new physics in this final state. The allowed parameter space region can be probed in the near future with low values for the significance. However, we optimized the current event selection criteria for new physics, finding out that the predicted particles could be discovered in the sensitivity range expected for the HL-LHC.
\newline
On the other hand, there are several projections for this work.  Among them, we can mention the need for studying and constraining the couplings between the new particles and SM leptons, because they are a source for lepton flavor violation. For instance, the model predicts the loop induced muon decay $\mu\to e\gamma$, which is highly constrained by experimental measurements. However, the loop calculation in this process is subtle, because the model presents a non minimal gauge structure. The non minimal gauge couplings are free parameters that modify the Feynman rules. Consequently, the radiative processes are highly sensitive to them and must be studied in depth.
\newline
Another interesting possibility is to consider the model in the context of dark matter phenomenology. The model presents two dark matter candidates, depending on the kinematical regime. If $M_N<M_{V^+}$, DM can annihilate only into SM leptons, and the relevant parameters for DM phenomenology are only $M_N$, the new vector masses and the $\beta$ couplings. On the other side, if $M_N>M_{V^+}$ the electroweak interactions are also relevant for DM studies, and therefore the relevant parameter space is more complex.
In both cases, the interaction between DM and leptons can be used to probe the model in the context of high energy cosmic rays, combining the constraints coming from collider physics and astrophysical sources. 



\backmatter

\bibliographystyle{utphys}
\bibliography{References}
\addcontentsline{toc}{chapter}{\bibname}

\end{document}